\documentclass[aip,reprint]{revtex4-1}

\draft 
\usepackage{graphicx, subfigure}
\usepackage{dcolumn}
\usepackage{bm}
\usepackage{ amsmath,amssymb, changepage}
\usepackage{rotating}
\usepackage{setspace}
\usepackage{enumitem, hyperref}
\usepackage{mathtools}
\usepackage{float}
\usepackage[bottom]{footmisc}
\usepackage{array}                      
\usepackage{MnSymbol}           
\usepackage[nice]{nicefrac}

\allowdisplaybreaks

\begin{document}
\title{Modelling the effect of acoustic waves on nucleation}

\author{S. R. Haqshenas}
\email[] {seyyed.haqshenas.12@ucl.ac.uk}
\affiliation{Department of Mechanical Engineering, University College London, Gower Street, London, WC1E 7JE, United Kingdom}
\author{I. J. Ford}
\email[]{i.ford@ucl.ac.uk} 
\affiliation{Department of Physics and Astronomy, University College London, Gower Street, London, WC1E 6BT, United Kingdom}
\author{N. Saffari}
\email[]{n.saffari@ucl.ac.uk} 
\affiliation{Department of Mechanical Engineering, University College London, Gower Street, London, WC1E 7JE, United Kingdom}

\date{14 July 2016}

\begin{abstract}
A phase transformation in a metastable phase can be affected when it is subjected to a high intensity ultrasound wave. In this study we determined the effect of oscillation in pressure and temperature on a phase transformation using the Gibbs droplet model in a generic format. The developed model is valid for both equilibrium and non-equilibrium clusters formed through a stationary or non-stationary process. We validated the underlying model by comparing the predicted kinetics of water droplet formation from the gas phase against experimental data in the absence of ultrasound. Our results demonstrated better agreement with experimental data in comparison with classical nucleation theory. Then, we determined the thermodynamics and kinetics of nucleation and the early stage of growth of clusters in an isothermal sonocrystallisation process. This new contribution shows that the effect of pressure on the kinetics of nucleation is cluster size-dependent in contrast to classical nucleation theory.
\end{abstract}


\maketitle 

\section{Introduction}\label{sec:intro}

A phase transformation in a liquid phase can be affected when it is subject to high intensity acoustic waves. The rarefaction pressure swing of the wave can nucleate bubbles or induce a liquid-gas transition, \cite{neppiras1951,blander1975, akulichev1982, baidakov1981} called acoustic cavitation. Several theoretical and experimental studies have shown that wave propagation in melts and supercooled liquids causes periodic phase transformation resulting in enhanced crystallisation. \cite{kapustin1963, akulichev1983,  arakelyan1987} For instance, liquid helium exposed to a high intensity focused ultrasound field undergoes liquid-solid transformation. \cite{chavanne2001} Nucleation of solid helium was observed to happen over the compression cycle followed by a decay and finally melting during the rarefaction cycle. Likewise, there is a body of works, mainly experimental, reporting the effect of an ultrasound field on crystallisation in a supersaturated solution. \cite{hem1967} The latter is usually referred as the sonocrystallisation process. Depending on the acoustic pressure magnitude and frequency, sonocrystallisation can yield a high nucleation rate and produce much finer crystals with a narrower crystal size distribution (CSD) compared to the conventional (silent) cooling crystallisation. \cite{hem1967, ruecroft2005}

The  mechanism by which ultrasound affects nucleation in a supersaturated solution is uncertain but it has been mainly attributed to the cavitation phenomenon. \cite{suslick1988} Enhancement in nucleation rate was, however, experimentally observed \cite{mazhul1954} in an ultrasound field that is weak enough to inhibit cavitation. Two main types of ultrasound-induced cavitation are inertial and stable cavitation. Inertial cavitation is the event when tiny cavities or dissolved gases in the liquid grow rapidly due to the rarefaction created by the ultrasound wave and collapse violently in the compression cycle of the ultrasound wave. This collapse generates enormous shock waves travelling with a speed of about $4000~\mathrm{ms^{-1}}$ and a magnitude of up to $1~\mathrm{GPa}$ as well as a temperature rise at the centre of the bubble to about $5000~\mathrm{K}$. \cite{flint1991, suslick2008} This can also lead to a significant temperature variation at a rate of $10^{10}~\mathrm{Ks^{-1}}$. \cite{flint1991, suslick2008} All these effects happen locally and over a very short period of time, i.e. spatially and temporally on scales of the volume of a bubble and nano-seconds respectively. \cite{akhatov2001, ohl1999} In the case of an asymmetric collapse, e.g. due to an oscillation and implosion of a bubble in the vicinity of a solid surface, a jet of fluid, at speeds greater than $100~\mathrm{ms^{-1}}$, is generated which can also influence the crystallisation process. \cite{suslick1999} Both the direct acoustic field and the indirect effects associated with cavitation influence the thermodynamics and kinetics of nucleation.

Considering the effect of static pressure on nucleation, Ford \cite{ford1991} modelled the pressure dependent homogeneous nucleation in a gas mixture using a statistical mechanics approach. Within the framework of classical nucleation theory (CNT), Kashchiev et al. \cite{kashchiev1995} proposed a model estimating the pressure dependent nucleation rate of condensed phase in a solution. This model does not consider the effect of pressure on the excess free energy as it was based on a cluster boundary defined by the equimolar dividing surface (EDS).  With regard to modelling the effect of acoustic cavitation on  crystallisation or solidification, the influence of radiated pressure from a collapsing bubble on the thermodynamics of ice formation was studied by Saclier et al. \cite{saclier2010} Louisnard et al. \cite{louisnard2007} , however, proposed a  segregation hypothesis where mass transportation due to the emitted shock wave from an inertial cavitation is the main factor leading to high nucleation rates observed experimentally. They suggested the mass diffusion mechanism and its effect on the kinetics of nucleation as the key factor rather than  the effect of pressure oscillation on the thermodynamic state. Here we show that this may only be the case if a cluster is defined by an EDS. Nevertheless, pressure fluctuation affects both the nucleation work and kinetics simultaneously.

If we model the kinetics of nucleation with the cluster dynamics approach, i.e. the master equation, \cite{kashchiev2000} it is determined by means of aggregative and non-aggregative mechanisms. Aggregative mechanisms include nucleation, growth and ageing, that give rise to the flux of cluster concentration along the size axis $n$. The non-aggregative mechanism accounts for change in composition (concentration of clusters) driven by mass flux along the space parameter axis. These two fluxes together determine the cluster distribution over time. We show that an acoustic wave can affect both processes which creates a coupled problem. However, depending on the magnitude and wavelength of the pressure fluctuation, the non-aggregative process, i.e. mass flux due to pressure gradient across space between adjacent systems within the bath, might be negligible. \cite{hirschfelder1954,bird1960}  This work aims to investigate the effect of pressure fluctuation on aggregative mechanism in particular. This allows us to study the effect of pressure variation on nucleation, the early stage of growth and also the Ostwald-ripening phenomenon. Furthermore, focusing on the aggregative mechanism we only need to know the local pressure fluctuation in the region of interest, i.e. dynamic pressure in the system, which can be emitted from any type of acoustic source, e.g. a planar or focused transducer or radiated pressure from either stable or inertial cavitation. Nevertheless, the development accounts for the effects associated with the wave propagation including temperature change, e.g. due to absorption, too. This makes it possible to apply this formulation to investigate the effect of pressure fluctuation in the old phase, emitted from any acoustic source, on the thermodynamics and kinetics of a first order phase change. Accounting for the non-aggregative effect of an acoustic wave and combining it with the present work to resolve the coupled problem is the subject of a forthcoming paper.

To accomplish our objectives, we use the Gibbs droplet model in a generic format to estimate the clustering work for both equilibrium and non-equilibrium clusters (Sec. \ref{sec:WorkCluster}). We develop equations to determine the number of molecules in both the new phase core, and the surface of a cluster defined by a non-EDS (Sec. \ref{sec:Nsigma}). We then demonstrate the significance of these improvements in reproducing the excess free energy of clusters obtained from statistical mechanics simulations and consequently estimating a nucleation rate by employing experimental data of water droplet formation from the vapour phase (Sec. \ref{sec:WaterDropletResults}). Using this model, we then show that the effect of pressure on kinetics is size-dependent and sensitive to the placement of dividing surface, especially for small clusters (Sec. \ref{sec:TranNucl}). It was previously shown \cite{ford2001, kashchiev2006} that the isothermal effect of pressure on nucleation work depends on the excess number of molecules in a nucleus and therefore is size-dependent too. We finally report in Sec. \ref{sec:ResultDisc} the effect of magnitude and frequency of acoustic waves on nucleation work and kinetics in an aqueous solution if we use a non-EDS cluster and compare it with results predicted by the CNT.

\section{Work of cluster formation}\label{sec:WorkCluster}

 Mechanical work is required to convert the old phase into the metastable state and start the formation of a new phase. The new phase is characterised as a cluster of molecules with a density that differs from the mother phase. This work becomes maximum in the case of formation of a critical cluster which is the cluster in unstable thermodynamic equilibrium with the old phase.

 Cluster formation work depends on the thermodynamic state and constraints applied to the old phase. We consider the system as a volume element coupled to a heat and particle bath. The phase change takes place within this system. The choice of heat and particle bath essentially means that temperature and volume of system remain constant and the old phase in the system has the same chemical potential as of the bath. This set of constraints is usually experimentally favoured and will be adopted in the following analysis. The system initially consists of the homogeneous old phase. After cluster formation, the system includes three phases, namely the core of cluster taking the new phase, the old phase surrounding the new phase and an interface phase. The new phase is considered as a homogeneous closed phase. The interface phase lies on an arbitrary dividing surface between the new and old phase considered as a Gibbs geometrical surface, i.e. a zero volume layer. The properties of the old phase are displayed below with no suffix whereas the suffices $n~\text{and}~\sigma$ label the new and interface phases, respectively.

 The reversible work of creating a cluster is equal to the change in the free energy of the system and is given by \cite{ford1996, ford2001,kashchiev2000}

\begin{equation}
\label{eq:Work1}
\begin{split}
\Delta \Omega(n_n,n_\sigma,p,T,x)~=&-n_n\Delta \mu(p,T,x) + \left( p - p_n  \right) V_n  \\
&- n_\sigma \Delta \mu_\sigma(p,T) + \int\limits_{p}^{p_n}V_ndp  +  \Omega_\sigma,
\end{split}
\end{equation}

\noindent where $\Delta \mu (p,T,x)$ is a difference in chemical potentials of old and new phases at temperature $T$, pressure $p$ and composition $x$ of the old phase, $n_n$ and $n_\sigma$ are the number of molecules in the new and interface phase, respectively, and $V_n$ is the cluster volume. $\Omega$ is the thermodynamic grand potential. Likewise, $\Omega_\sigma$ is the grand potential associated with the interface phase which is also represented by $A_\sigma \gamma$ where $ A_\sigma$ is the interfacial surface area and $\gamma$ is the surface tension. The size of cluster (in molecules) is equal to $n=n_n+n_\sigma$ and its volume (for a single component cluster) is given by $V_n=n_n\nu_n$ where $\nu_n=\tfrac{1}{\rho_n}$ is the specific volume of the new phase. The difference between the chemical potential of the old phase and the bulk new phase evaluated at the temperature and pressure of the old phase reads

\begin{equation}
\label{eq:ChemPot1}
\Delta \mu (p,T,x)~=~\mu(p,T,x)~-~\mu_n(p,T),
\end{equation}
\noindent and similarly, $\Delta \mu_\sigma(p,T,x) =\mu(p,T,x) -\mu_\sigma(p,T)$.  For the sake of briefness in the notation, independent variables, i.e. $p,T,x$, will not be displayed unless it is required. Nevertheless, we shall note that they could vary over time in the system and time and space in the bath. In the case of a condensed new phase, the cluster can be considered practically incompressible. Thus, the work of formation of a condensed cluster becomes

\begin{equation}
\label{eq:WorkCond}
\Delta \Omega= -n_n\Delta \mu  - n_\sigma \Delta \mu_\sigma +  \Omega_\sigma,
\end{equation}

\noindent and substituting $ \Omega_\sigma$ with $ A_\sigma \gamma$ and rearranging the above equation gives

\begin{equation}
\label{eq:WorkCond2}
 \Delta \Omega= -(n_n + n_\sigma)\mu+ n_n \mu_n + n_\sigma\mu_\sigma + A_\sigma \gamma.
\end{equation}

To be able to use these equations, $ \mu _\sigma$ and $n_\sigma$ should also be determined for a generic dividing surface. This is discussed in Secs. \ref{sec:Nsigma}-\ref{sec:NonCriticalCLuster}.

\subsubsection{Number of surface molecules} \label{sec:Nsigma}
\begin{figure}
\centering
\includegraphics{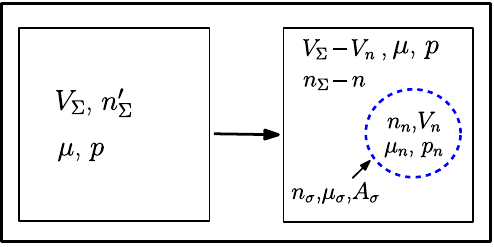}
\caption{Cluster formation in a system with constant volume, temperature and chemical potential. $n^\prime_\Sigma$ and $n_\Sigma$ are the number of molecules in the system before and after cluster formation, respectively. Refer to the text for details.}
\label{fig:Bath_system}
\end{figure}

 For the system containing a new cluster within the old phase, we can write \cite{guggenheim1985}

\begin{equation}
\label{eq:InterfaceNo}
n_\Sigma = \rho_n V_n + \rho (V_\Sigma - V_n) + n_\sigma ,
\end{equation}

\noindent where $\Sigma$ represents the entire system including all three phases, and $\rho_n$ and $\rho$ are the molecular number density of the new and old phase, respectively (see Figure \ref{fig:Bath_system}). This equation can be re-arranged to $n_\Sigma - \rho V_\Sigma = (\rho_n  - \rho) V_n + n_\sigma $  where the left hand side (LHS) is invariant with respect to the choice of the dividing surface. If we choose an EDS, we will have by definition $n_\sigma=0$ \cite{kashchiev2000} and therefore $\mathrm{LHS}=(\rho_n  - \rho) V_n^{E} $ where $V_n^{E}$ is the volume of a cluster defined using the EDS. Given that LHS is invariant with the choice of surface, the equation for an arbitrary surface becomes $\mathrm{LHS}=(\rho_n  - \rho) V_n^{E}=(\rho_n  - \rho) V_n + n_\sigma $ which yields

\begin{equation}
\label{eq:InterfaceNoFinal}
n_\sigma=(\rho_n  - \rho) (V_n^{E} - V_n),
\end{equation}

\noindent and substituting $V_n=n_n \nu_n=\tfrac{n_n}{\rho_n}$ and $V_n^E=n_e\nu_n=\tfrac{n_e}{\rho_n}$ where $n_e$ is the size of an EDS-defined cluster, this equation simplifies to

\begin{equation}
\label{eq:InterfaceNoFinal2}
n_\sigma = (1  - \dfrac{\rho}{\rho_n}) (n_e - n_n) ~=~k_\rho (n_e - n_n),
\end{equation}

\noindent where $k_\rho = 1  - \tfrac{\rho}{\rho_n}$. This is a generic equation and valid for any shape of cluster. We can write

\begin{equation}
\label{eq:nExcess}
k_\rho n_e ~=~ k_\rho n_n + n_\sigma ~=~\Delta n_{exc} ,
\end{equation}

\noindent where $\Delta n_{exc}$ is the excess number of molecules in the cluster of volume $V_n$ comparing to the same volume of old phase. This quantity is independent of the choice of dividing surface. In the case of a condensed new phase we have $k_\rho>0$ and consequently $\Delta n_{exc}>0$. However, when the new phase is less denser than the old phase, e.g. bubble formation,  $k_\rho$ and $\Delta n_{exc}$ become negative.

As shown in Appendix \ref{sec:AppxNsigma}, for cubic and spherical clusters we have $n_\sigma = k_\rho \mathcal{G}(n_n)$ with

\begin{eqnarray}
\label{eq:Nsigma}
\mathcal{G}(n_n) =   \dfrac{3}{S_f} \lambda \left(    n_n^{\scriptscriptstyle{\tfrac{2}{3}} }  +  \dfrac{\lambda}{S_f}  n_n^{\scriptscriptstyle{\tfrac{1}{3}} } + \dfrac{\lambda^2}{3 S_f^2}   \right),
\end{eqnarray}

 \noindent where  $\lambda=\tfrac{\delta}{R_0}$ is a dimensionless quantity that distinguishes an arbitrary dividing surface from the EDS. Here $\delta=R_\sigma^{E}-R_\sigma$ is the radial separation between the EDS and the arbitrary surface and $R_0=\left(\tfrac{3\nu_n}{4 \pi}\right)^{\scriptscriptstyle {1/3}}$ is the radius of a molecule in the new phase, considered to be a sphere. $S_f$ is the shape factor which equals to unity for a spherical cluster. Henceforth, we assume the cluster is spherical.  The total size of a cluster then reads $n=n_n+n_\sigma = n_n+k_\rho \mathcal{G}(n_n)$. Depending on the density of new and old phases and the location of the dividing surface, $n_\sigma$ can become positive or negative. This model satisfies the following conditions

\begin{equation}
\label{eq:InterfaceLim}
\lim\limits_{n\rightarrow \infty} \dfrac{n_\sigma}{n}=0 \hspace{5mm} \mathrm{,} \hspace{5mm}  \lim\limits_{n\rightarrow \infty} \dfrac{n_n}{n}=1.
\end{equation}

\noindent They imply that for large clusters the number of molecules in the core becomes dominant and the EDS becomes acceptable for defining the boundary of a cluster.  However, for a small cluster for which a core with bulk properties does not exist, the contribution of interface phase takes on an important role which can be modelled through interface terms with non-zero $n_\sigma$ ($\lambda$ in our model).

If the arbitrary surface is selected such that it coincides with the surface of tension, then $R_\sigma=R_t$ and in the limits of $R_t \rightarrow \infty$ the separation length converges to the Tolman length $\delta \rightarrow \delta_T$ and subsequently $\lambda \rightarrow \lambda_T=\tfrac{\delta_T}{R_0}$.

Now we need to obtain $\mu_\sigma$ which depends on the condition of a cluster. This is addressed in the Secs. \ref{sec:CriticalCLuster} and \ref{sec:NonCriticalCLuster}.

\subsubsection{Critical cluster} \label{sec:CriticalCLuster}

The critical cluster is in an unstable thermodynamic equilibrium with the old phase and satisfies the following conditions:\cite{abraham1974, ford2001} $\mu_n^*(p_n^*)=\mu_\sigma^*=\mu$, and the well-known Laplace equation $p_n^*-p=\tfrac{d\Omega^*_\sigma}{dV_n^*}$ where the asterisk denotes the properties of the critical cluster. Substituting these relations in Eq. \ref{eq:WorkCond} gives the work of formation of the critical cluster as follows

\begin{eqnarray}
\label{eq:CriticalWork}
\Delta \Omega^*&=&-n_n^*\Delta \mu + \Omega_\sigma^* = -(p_n^* - p)V_n^* + \Omega_\sigma^*  \nonumber \\
&=& -\dfrac{d\Omega^*_\sigma}{dV_n^*}V_n^*  + \Omega_\sigma^* = -n_n^* \dfrac{d\Omega_\sigma^*}{dn_n^*}   + \Omega_\sigma^*,
\end{eqnarray}

\noindent given that $n^*=n_n^*+n_\sigma^*$, this equation may be reformulated as

\begin{equation}
\label{eq:CriticalWorkForm2}
\Delta \Omega^*=-n^*\Delta \mu + n_\sigma^*\Delta \mu + \Omega_\sigma^*.
\end{equation}

\noindent The last two terms in the above equation essentially represent the excess Helmholtz free energy of the interface phase of the critical cluster size $n^*$, i.e. $F_s^* ~=~\Omega_\sigma^* + n_\sigma^* \Delta \mu $. \cite{ford1996}

The grand potential of the interface phase can be written as a function of $n_\sigma$ \cite{kalikmanov2012} or the area of the cluster, basically a function of $n_n$. In any case, we can plausibly consider $\Omega_\sigma=\Omega_\sigma(n_n)$. Therefore, the Taylor series expansion of $\Omega_\sigma$ about $n$ reads

\begin{eqnarray}
\label{eq:OmegaTaylor}
\Omega_\sigma(n_n) ~&=& \Omega_\sigma(n) - n_\sigma \dfrac{d\Omega_\sigma}{d n_n} \bigg|_n  + \dfrac{n_\sigma^2}{2} \dfrac{d^2 \Omega_\sigma}{d n_n^2}\bigg|_n  \nonumber \\
&&- \dfrac{n_\sigma^3}{6} \dfrac{d^3 \Omega_\sigma}{d n_n^3}\bigg|_n + \mathcal{O}(n_\sigma^4).
\end{eqnarray}

\noindent Evaluating this equation at the critical cluster and inserting the results in the first formula of Eq. \ref{eq:CriticalWork} gives

\begin{eqnarray}
\label{eq:CriticalWork2}
\Delta \Omega^*&=& \Omega_\sigma(n^*)  - \left( n_n^*\dfrac{d\Omega_\sigma}{dn_n}\bigg|_{n_n^*} + n_\sigma^*\dfrac{d\Omega_\sigma}{dn_n}\bigg|_{n^*} \right) \nonumber \\
&&  + \dfrac{{n_\sigma^*}^2}{2} \dfrac{d^2 \Omega_\sigma}{d {n_n}^2} \bigg|_{n^*} - \dfrac{{n_\sigma^*}^3}{6} \dfrac{d^3 \Omega_\sigma}{d {n_n}^3}\bigg|_{n^*} + \mathcal{O}(n_\sigma^4).
\end{eqnarray}

\noindent If a cluster is defined by the EDS and the capillarity approximation is imposed, the above equation simplifies to the nucleation work given by the CNT.

\subsubsection{Non-critical cluster}\label{sec:NonCriticalCLuster}

The equality of chemical potentials of all phases may not hold for a non-critical cluster, i.e. a non-equilibrium cluster, which makes it a complicated situation to analyse. For a non-critical cluster we assume  $\mu_\sigma \approx \mu_n(p_n)$. This assumption is justified if diffusive exchange of molecules from interface phase to the new phase is faster than the diffusion of molecules towards the interface from the old phase. \cite{vehkamaki2006} This gives  $\Delta \mu _\sigma \approx \Delta \mu - \nu_n(p_n-p)$ and subsequently Eq. \ref{eq:WorkCond} transforms to

\begin{equation}
\label{eq:WorkFinalDelP}
\Delta \Omega= -n\Delta \mu  + n_\sigma \nu_n (p_n-p) +  \Omega_\sigma.
\end{equation}

\noindent Now we need to determine the quantity $p_n-p$ for a non-critical cluster. This is not a trivial problem and needs knowledge from statistical or molecular models. Nevertheless, the following two methods have previously been used to estimate this quantity using continuum thermodynamics.\cite{vehkamaki2006} In the first method, we use the Maxwell relationship of $d\mu_n=\nu_ndp_n$, under isothermal conditions, and obtain the exact equation $p_n-p=\rho_n\Delta \mu + (p_n-p_n^*)$: see Appendix \ref{sec:AppxChemPot} for the derivation. The last term is the difference between the inner pressure of a non-critical cluster and a critical cluster for the same pressure of the old phase $p$. Inserting this into Eq. \ref{eq:WorkFinalDelP} gives

\begin{equation}
\label{eq:WorkFinalDelPV1}
\Delta \Omega= -n\Delta \mu  + n_\sigma \left( \Delta \mu + \nu_n (p_n-p_n^*) \right) +  \Omega_\sigma.
\end{equation}

\noindent Given that $\nu_n$ is in the order of few $\mathrm{nm^3}$, we can approximate the second term in the above equation by $n_\sigma \Delta \mu$ as $\nu_n (p_n-p_n^*)$ is relatively small comparing to $\Delta\mu$. Therefore this equation simplifies to

\begin{equation}
\label{eq:WorkFinalDelPV1Appr0}
\Delta \Omega= -n\Delta \mu  + n_\sigma \Delta \mu  +  \Omega_\sigma  = -n_n\Delta \mu  +  \Omega_\sigma ,
\end{equation}

 \noindent which can also be written as

 \begin{equation}
 \label{eq:WorkFinalDelPV1Appr}
 \Delta \Omega= -n\Delta \mu  + F_{s,1},
 \end{equation}

\noindent where $F_{s,1} ~=~ n_\sigma \Delta \mu+\Omega_\sigma$. Making a comparison between Eq. \ref{eq:WorkFinalDelPV1Appr0} and Eq. \ref{eq:WorkCond}  reveals that this approximation essentially sets $\Delta \mu_\sigma(p,T)$ to zero. This condition is a result of mathematically cancelling the pressure term against the supersaturation term in the former equations while not enforcing the physical equilibrium conditions. Evaluating this equation for the critical cluster $n^*$ yields Eq. \ref{eq:CriticalWorkForm2} as anticipated. This tells that the work of formation of a non-critical cluster can be reasonably approximated by the equation that determines the work of formation of a critical cluster.

In the second method, the pressure difference between the inside and outside of a cluster is approximated using the generalised Laplace equation $p_n-p=\tfrac{d\Omega_\sigma}{dV_n}$. \cite{kashchiev2000} This method basically assumes that Laplace equation could be extended to sub-critical and supercritical clusters. Employing this approximation transforms Eq. \ref{eq:WorkFinalDelP} to

\begin{equation}
\label{eq:WorkFinal}
\Delta \Omega= -n\Delta \mu  + n_\sigma \nu_n \dfrac{d\Omega_\sigma}{dV_n} +  \Omega_\sigma,
\end{equation}

\noindent and given $\tfrac{d\Omega_\sigma}{dV_n}=\tfrac{d\Omega_\sigma}{dn_n}.\tfrac{d n_n}{dV_n}= \tfrac{d\Omega_\sigma}{dn_n}.\tfrac{1}{\nu_n}$ it follows that

\begin{equation}
\label{eq:WorkFinalDelPV2Nn}
\Delta \Omega= -n\Delta \mu  + n_\sigma \dfrac{d\Omega_\sigma}{dn_n} +  \Omega_\sigma  = -n\Delta \mu  + F_{s,2},
\end{equation}

\noindent where $F_{s,2} ~=~ n_\sigma \dfrac{d\Omega_\sigma}{dn_n} +  \Omega_\sigma $. This equation simplifies to the following relationship by using Eq. \ref{eq:OmegaTaylor}

\begin{eqnarray}
\label{eq:WorkFinalDelPV2N}
\Delta \Omega ~ &=& -n\Delta \mu  + \Omega_\sigma(n) + n_\sigma \left( \dfrac{d\Omega_\sigma}{dn_n} - \dfrac{d\Omega_\sigma}{dn_n}\bigg|_{n}  \right) \nonumber \\
&& + \dfrac{n_\sigma^2}{2} \dfrac{d^2 \Omega_\sigma}{d n_n^2}\bigg|_{n} - \dfrac{n_\sigma^3}{6} \dfrac{d^3 \Omega_\sigma}{d n_n^3}\bigg|_{n} + \mathcal{O}(n_\sigma^4).
\end{eqnarray}

Both Eqs. \ref{eq:WorkFinalDelPV1Appr} and \ref{eq:WorkFinalDelPV2Nn} give an approximation of the cluster formation work over the entire range of cluster size. They become identical for an EDS cluster. However, evaluating Eq. \ref{eq:WorkFinalDelPV2Nn} for the critical non-EDS cluster gives a nucleation work which is different from the nucleation work obtained from the exact Eq. \ref{eq:CriticalWork2}. Since we are interested in a formulation that estimates cluster formation work for both critical and non-critical non-EDS clusters, Eq. \ref{eq:WorkFinalDelPV1Appr} suits our needs better and will be utilised in this work. It should be noted that nucleation work is a physical property of the system and is independent of the location of dividing surface. Consequently, the desired formulation has to agree with the result of the exact Eq. \ref{eq:CriticalWork2} for the critical cluster with non-zero $n_\sigma$.

In spite of this, the work of formation of a cluster varies depending on the way it is being identified. To elaborate on this, we keep the cluster size $n$ constant and compare the work of formation of a classical cluster (identified with EDS and capillarity approximation) with a generic cluster of the same size. The identical cluster size implies equivalent ``bulk'' work ( i.e. $-n\Delta \mu$, note that this is different from volume work as volume depends on $n_n$ not $n$) whereas the excess free energy is different. The ratio of excess free energies is determined by

 \begin{equation}
 \label{eq:FsratioGen}
 F_{s,r}=\dfrac{F_{s,1}}{F_{s,cl}}=\dfrac{n_\sigma \Delta \mu  +  \Omega_\sigma}{\Omega_{\sigma,cl}}.
 \end{equation}

 \noindent Considering $\Omega_\sigma = A_\sigma \gamma(n)=a_{\scriptscriptstyle 0} \gamma(n) n_n^{ \nicefrac{2}{3}}$ where $a_{\scriptscriptstyle 0}=4 \pi R_0^2$ and $\Omega_{\sigma,cl} = A_{cl} \gamma_{\scriptscriptstyle \infty} = a_{\scriptscriptstyle 0} \gamma_{\scriptscriptstyle \infty} n^{\nicefrac{2}{3}}$ where $ \gamma_{\scriptscriptstyle \infty}$ is the planar surface tension between two phases in equilibrium, this equation simplifies to

 \begin{eqnarray}
 \label{eq:FsratioSimp}
 F_{s,r} ~&=&\dfrac{n_\sigma \Delta \mu  +  a_{\scriptscriptstyle 0} \gamma(n) n_n^{\tfrac{2}{3}}}{a_{\scriptscriptstyle 0} \gamma_{\scriptscriptstyle \infty} n^{\tfrac{2}{3}}} \nonumber \\
 &=& \dfrac{ \Delta \mu }{a_{\scriptscriptstyle 0} \gamma_{\scriptscriptstyle \infty} } \left( \dfrac{n_\sigma}{n^{\tfrac{2}{3}}} \right) +  \dfrac{\gamma(n)}{\gamma_{\scriptscriptstyle \infty} } \left(\dfrac{ n_n}{n}\right)^{\tfrac{2}{3}}.
 \end{eqnarray}

 \noindent  We can use an EDS cluster to define the effective surface free energy as $ F_{s,\mathrm{\scriptscriptstyle eff}} = \Omega_{\sigma,\mathrm{\scriptscriptstyle eff}} = A_{\scriptscriptstyle cl} \gamma_{\mathrm{\scriptscriptstyle eff}} $. Considering that the majority of simulations conducted by means of molecular dynamics (MD) or other statistical mechanical approaches report results for EDS defined clusters, this choice also allows us to make a comparison between our $\gamma_{\mathrm{\scriptscriptstyle eff}}$ and excess free energy with their counterparts in those works. Setting $F_{s,1} = F_{s,\mathrm{\scriptscriptstyle eff}} $ yields $ \gamma_{\mathrm{\scriptscriptstyle eff}} = \gamma_{\scriptscriptstyle \infty} F_{s,r} $. Substituting $F_{s,r} $ from Eq. \ref{eq:FsratioSimp} gives

 \begin{equation}
 \label{eq:GammaEff}
 \gamma_{\mathrm{\scriptscriptstyle eff}} = \dfrac{ \Delta \mu }{a_0 } \left( \dfrac{n_\sigma}{n^{\tfrac{2}{3}}} \right) +  \gamma(n) \left(\dfrac{ n_n}{n}\right)^{\tfrac{2}{3}}.
 \end{equation}

The first term accounts for temperature and concentration dependence of the effective surface tension and the second term describes the curvature dependence of the effective surface tension. The fact that concentration influences the excess free energy and consequently the effective surface tension, according to our definition, was demonstrated and formulated in different works too. \cite{baidakov1997, ford1997}

Now we specify the dividing surface and associated surface tension which will be used in this work. Given Eq. \ref{eq:Nsigma} we can use any arbitrary surface which can more precisely approximate the excess free energy. Utilising the EDS ($\lambda=0$) to define a cluster along with the capillarity approximation yields the conventional form of the CNT which is unsuccessful in explaining many experimental observations. This can be improved if the size dependency of surface tension is accounted for. If the cluster's boundary is identified by the surface of tension, the curvature dependence of the surface tension could be accounted for by the Tolman equation\cite{tolman1949} or other polynomial expansion models.\cite{helfrich1973} It has been shown that by choosing the surface of tension ($\lambda=\lambda_T$) and employing the Tolman equation to correct the surface tension,  we can achieve a better agreement with some experiments. \cite{wilhelmsen2015} MD or DFT simulations are usually required to determine an appropriate Tolman length.

Nevertheless, the Tolman equation is useful for larger clusters but is expected to break down for small clusters. Therefore, we opt not to employ it in this work, instead we define a non-EDS surface identified by the parameter $\lambda$, and the size-independent surface tension of $\gamma_\infty$ (denoted as \emph{the new surface} in this work). This leads to $\gamma(n)=\gamma_\infty$ and consequently $\Omega_\sigma = a_{\scriptscriptstyle 0} \gamma_\infty n_n^{\nicefrac{2}{3}}$. Fitting $\gamma_{\mathrm{\scriptscriptstyle eff}}$ of Eq. \ref{eq:GammaEff} to the effective surface tension obtained from statistical mechanical methods yields the value of $\lambda$. In the case of nucleation of water droplets discussed in Sec. \ref{sec:WaterDropletResults}, $\lambda$ is estimated with the aid of excess free energies of clusters obtained from statistical mechanics simulations using the TIP4P/2005 molecular model. For the crystallisation exercise, however, we performed simulations at different values of $\lambda$ to investigate its influence on the kinetics of crystallisation under pressure fluctuation. We should point out that Kashchiev \cite{kashchiev2003} introduced a non-equimolar dividing surface identified by the condition $\tfrac{d \gamma(n)}{d n}=0$, termed the conservative surface. For the conservative surface, the surface tension is also size-independent and equals the macroscopic planar value $\gamma_\infty$. The question of whether the new surface adopted in this work coincides with the conservative surface is out of the scope of the current paper and will be addressed elsewhere.

Having determined the specifications of the dividing surface, the excess number of molecules and clustering work, we can study the process of cluster formation and determine the size of the nucleus and nucleation work when the old phase is exposed to an acoustic wave. Furthermore, the kinetics of nucleation under this circumstance should be calculated. This is discussed in Secs. \ref{sec:PertEffect} and \ref{sec:TranNucl}.

\section{Effect of perturbation in old phase on the clustering work} \label{sec:PertEffect}

\noindent As the acoustic wave propagates in the bath, it causes pressure fluctuation, temperature variation and mass transportation due to a spatial pressure gradient. \cite{bird1960} The effect of variation of pressure, temperature and composition on the work of formation of a cluster then should be evaluated. The total differential of $\Delta \Omega$ is calculated by differentiating Eq. \ref{eq:WorkCond2} which gives

\begin{eqnarray}
\label{eq:WorkCondDiffTot}
d\Delta \Omega~ & =& -(n_n + n_\sigma)d\mu -(dn_n + dn_\sigma)\mu + n_n d\mu_n + dn_n \mu_n  \nonumber \\
 &&  +n_\sigma d\mu_\sigma + dn_\sigma \mu_\sigma + dA_\sigma \gamma + A_\sigma d\gamma,
\end{eqnarray}

\noindent  and subsequently, this re-arranges to

\begin{eqnarray}
\label{eq:WorkCondDiffTot2}
 d\Delta \Omega ~ &=& -(n_n + n_\sigma)d\mu + n_n d\mu_n + n_\sigma d\mu_\sigma + A_\sigma d\gamma    \nonumber \\
 &&    -dn_n( \mu  - \mu_n)  - dn_\sigma (\mu - \mu_\sigma) + dA_\sigma \gamma.
\end{eqnarray}

To compare with the silent condition, we are interested in evaluating the effect of acoustic wave on the work of formation of the same size cluster. This is obtained by setting $dn_n=dn_\sigma=dA_\sigma=0$ in the above equation which simplifies to

\begin{equation}
\label{eq:WorkCondDiff}
 d\Delta \Omega = -(n_n + n_\sigma)d\mu+ n_n d\mu_n + n_\sigma d\mu_\sigma + A_\sigma d\gamma.
\end{equation}

\noindent The change in the chemical potential of the old phases with respect to pressure and temperature can be estimated using a Gibbs-Duhem relation 

\begin{eqnarray}
\label{eq:ChemPot3}
d \mu=\left ( \dfrac{\partial \mu}{\partial T} \right )_{p}dT + \left ( \dfrac{\partial \mu}{\partial p} \right )_{T}dp = -sdT+\nu dp, 
\end{eqnarray}

\noindent where $\nu$ and $s$ are the partial molecular volume and entropy of the old phase. Likewise for the interface phase $d \mu_\sigma = -s_\sigma dT - \frac{A_\sigma}{n_\sigma} d\gamma$  and for the new phase $d \mu_n =-s_ndT + \nu_n dp$: see Appendix \ref{sec:AppxChemPot} for the derivation of the latter equation. Substituting $d\mu, d\mu_n$ and $d\mu_\sigma$ in the previous equation then gives

\begin{eqnarray}
\label{eq:WorkdiffCondFinal}
d\Delta \Omega~ &=& - \left[ -s(n_n+n_\sigma)+n_ns_n+n_\sigma s_\sigma \right] dT \nonumber \\
&& -\left[\nu(n_n+n_\sigma)-n_n \nu_n \right] dp,
\end{eqnarray}

\noindent which can be written as

\begin{equation}
\label{eq:WorkdiffCondFinal2}
d\Delta \Omega=- \Delta s_{exc}dT -\nu \Delta n_{exc}dp,
\end{equation}

\noindent where $\Delta s_{exc}=-s(n_n+n_\sigma)+n_ns_n+n_\sigma s_\sigma$ is the excess entropy gained by the system through the formation of a cluster of size $n$ and $ \Delta n_{exc}$ is defined in Eq. \ref{eq:nExcess}.

In the case of isothermal acoustic wave propagation, the effect of a pressure perturbation on the work of forming an $n$-sized cluster is determined by

\begin{equation}
\label{eq:WorkdiffCondPress}
\left( \dfrac{\partial \Delta \Omega}{\partial p}\right)_{T} = - \nu(n_n+n_\sigma) + n_n \nu_n = - \nu \Delta n_{exc}.
\end{equation}

\noindent Finally, integrating Eq. \ref{eq:WorkdiffCondFinal2} gives the work required to form clusters at a temperature, pressure and composition which differ from the reference state. This is expressed as

\begin{eqnarray}
\label{eq:WorkdiffCondFinalBetaInteg}
\Delta \Omega(n_n,n_\sigma,p,T)~ &=& \Delta \Omega_0(n_n,n_\sigma) -\int_{T_0}^T \Delta s_{exc} dT\nonumber \\
&&-~ \int_{p_0}^p\nu \Delta n_{exc} dp,
\end{eqnarray}

\noindent where $\Delta \Omega_0(n_n,n_\sigma)$ is the work required to create an $n$-sized cluster while the system is at the reference thermodynamic state ($T_0,p_0,x_0$). If the reference condition is chosen such that $x_0~=~x_e$, where $x_e$ is the equilibrium fractional concentration of monomers, then $\Delta \mu(p_0,T_0,x_0) = 0$ and $\Delta \Omega_0=-n_\sigma \Delta \mu_{\sigma,0} + A_\sigma \gamma$.

However, in sonocrystallisation experiments a supersaturated solution is usually made first, and then an acoustic wave is introduced. Therefore, it is practically desirable to choose the supersaturated state in silent condition, i.e. prior to application of acoustic wave, as the reference state. The difference in chemical potentials at the reference state needs then to be obtained. For the crystallisation process, we can write the partial molecular chemical potential of solute species in an ideal solution as \cite{guggenheim1985} $\mu~=~\mu^0~+~k_B T\mathrm{ln}(x)$ where $k_B$ is the Boltzmann constant. It is presumed that the solution is sufficiently dilute, such that the activity can be estimated by the concentration of solute molecules. $\mu^0$ is the chemical potential of the pure liquid at the same temperature and pressure and does not depend on the composition $x$. Consequently, at constant pressure and temperature of the reference state we can write $\Delta \mu_0 = k_B T_0 \mathrm{ln} (\frac{x_0}{x_e})$ where $x_e$  is evaluated at $T_0$ and $p_0$.

\subsection{Nucleation work}

Substituting the relationships describing the critical cluster in Sec. \ref{sec:CriticalCLuster} together with $\mu_n(p) = \mu_n(p_n)$ $- \nu_n \left( p_n - p \right)$ (see Appendix \ref{sec:AppxChemPot}) and $V_n=n_n \nu_n$, in Eq. \ref{eq:WorkCondDiffTot}, we can determine the variation in nucleation work due to change in properties of the old phase as follows

\begin{eqnarray}
\label{eq:NucWorkDiff}
d \Delta\Omega^*~& =&  -(n_n^* + n_\sigma^*)d\mu + n_n^* d\mu_n^* + n_\sigma^* d\mu^*_\sigma + A_\sigma^* d\gamma    \nonumber \\
 &&    -dn_n^*( \mu  -  \mu_n(p_n^*) + \nu_n \left( p_n^* - p \right)) \nonumber \\
 &&   - dn_\sigma^* (\mu - \mu_\sigma^*) + dA_\sigma^* \gamma,
\end{eqnarray}

\noindent which applying equilibrium conditions simplifies to

\begin{equation}
\label{eq:NucWorkDiff2}
d \Delta\Omega^* =  -(n_n^* + n_\sigma^*)d\mu + n_n^* d\mu_n^* + n_\sigma^* d\mu_\sigma^* + A_\sigma^* d\gamma.
\end{equation}

\noindent This equation is akin to Eq. \ref{eq:WorkCondDiff} being evaluated at the critical cluster size which reads

\begin{eqnarray}
\label{eq:NucWorkDiff3}
&& d\Delta \Omega^*~=~- \Delta s^*_{exc}dT -\nu \Delta n^*_{exc}dp,
\end{eqnarray}

\noindent where $\Delta s^*_{exc}$ and $ \Delta n^*_{exc}$ are the excess quantities evaluated for a critical cluster. Evaluating Eq. \ref{eq:WorkdiffCondFinalBetaInteg} at the size of a critical cluster at the reference condition gives the integral form of this equation.

\subsection{Nucleus size}

The fact that $\Delta \Omega>0$ implies that the work required for cluster formation becomes a maximum for a critical cluster. The size of critical cluster is then the extremum of the equilibrium equation which can be found by solving $\tfrac{d \Delta \Omega}{d n}=0$ for $n$. Differentiating Eq. \ref{eq:WorkdiffCondFinalBetaInteg} with respect to $n$  gives

\begin{eqnarray}
\label{eq:Work9}
&& \dfrac{\partial \Delta \Omega}{\partial n} =\dfrac{\partial \Delta \Omega_0}{\partial n}  - \int_{T_0}^T\Delta s^\prime_{exc} dT - \int_{p_0}^p \Delta n^\prime_{exc} \nu dp,
\end{eqnarray}

\noindent where $ \Delta s^\prime_{exc}=-s + s_n h^\prime (n_n) + s_\sigma h(n_n)$ and $\Delta n^\prime_{exc}=1 - \tfrac{\rho}{\rho_n} h^\prime (n_n)$ where $h(n_n)$ and $h^\prime(n_n)$ are derivatives of $n_\sigma$ and $n_n$ with respect to the cluster size $n$ and are given in Appendix \ref{sec:AppxNsigma}. Using Eq. \ref{eq:WorkFinalDelPV1Appr} for the reference state and identifying clusters with \emph{the new surface} yields

 \begin{eqnarray}
 \label{eq:Work11}
 \dfrac{\partial \Delta \Omega}{\partial n} ~&=& -k_B T_0 \mathrm{ln} (\dfrac{x_0}{x_e})(1-h(n_n)) + a_{\scriptscriptstyle 0} \gamma_{\scriptscriptstyle \infty}   \dfrac{2}{3} n_n^{-\tfrac{1}{3}}h^{'}(n_n) \nonumber \\
 &&  - \int_{T_0}^T\Delta s^\prime_{exc} dT - \int_{p_0}^p \Delta n^\prime_{exc} \nu dp.
 \end{eqnarray}

\noindent This is a complete equation for calculation of the variation of work required for cluster formation with respect to the size of cluster at different thermodynamic states. The size of the critical cluster is the root of $\tfrac{\partial \Delta \Omega}{\partial n} ~=~0$.

\noindent The generic solution for the critical cluster size in an arbitrary state depends on material properties and change in density and entropy of all phases with pressure and temperature, respectively. The solution for special cases though can be derived. The case of an isothermal process with incompressible old and new phases is considered and discussed below.

\subsection{Incompressible solution and isothermal condition} \label{sec:IsoThermNucl}

The absorption of propagating acoustic waves in a medium mainly depends on the viscosity of the medium and the wavelength. Wave propagation in an aqueous medium can be considered as an isothermal process since the absorption is low, especially during a short exposure. Additionally, if the partial molecular density of the old and new phases are pressure independent, Eq. \ref{eq:Work11} simplifies to

 \begin{eqnarray}
 \label{eq:Work11_Isotherm}
  \dfrac{\partial \Delta \Omega}{\partial n}~&=& -k_B T_0 \mathrm{ln} (\dfrac{x_0}{x_e})   ~+ ~a_{\scriptscriptstyle 0} \gamma_{\scriptscriptstyle \infty}   \dfrac{2}{3} n_n^{-\tfrac{1}{3}}h^{'}(n_n) \nonumber \\
 && +k_B T_0 \mathrm{ln} (\dfrac{x_0}{x_e}) h(n_n) - \Delta n^\prime_{exc} \nu \Delta p,
 \end{eqnarray}

\noindent where $\Delta p=p-p_0$ is the variation in pressure. Replacing $n$ with $n_n+k_\rho \mathcal{G}(n_n)$ and setting this relationship equal to zero gives a polynomial equation with $n_n$ unknown. We can numerically solve this equation and obtain $n_n^*$ and consequently $n^*$. If we define the cluster by EDS, this equation simplifies and gives the analytic solution for the size of nucleus ($n_e^*$) as follows

\begin{equation}
\label{eq:NucleiSizeSphereCNT}
n_e^*=\left( \dfrac{2}{3}~ \dfrac {\dfrac{a_0  \gamma_{\scriptscriptstyle \infty} }{k_B T_0} } { \mathrm{ln} (r_0)~+~\dfrac{k_\rho \nu \Delta p }{k_B T_0}  }  \right)^{3},
\end{equation}

\noindent where $r_0=\tfrac{x_0}{x_e}$. In the crystallisation process, $r_0$ is the ratio of solute molar concentration to the equilibrium molar concentration at the reference state at initial time instant.

\noindent This equation demonstrates the effect of pressure fluctuation and variation in composition on the size of EDS-defined nuclei. This equation for EDS clusters was first derived by Kashchiev and van Rosmalen. \cite{kashchiev1995}

Having determined $n^*$, the nucleation work is given by Eq. \ref{eq:WorkFinalDelPV1Appr} for the cluster size of $n^*$.

 We established the thermodynamics of equilibrium and non-equilibrium clusters based on the Gibbs droplet model with an arbitrary dividing surface. The conservation of mass was used to determine the number of molecules in the interface phase ($n_\sigma$) as a function of the cluster size $n$. We also calculated the effective surface tension of this arbitrary surface and demonstrated its size and chemical potential dependencies. The new development may resemble a classical model with a variable surface tension as a function of the cluster size and chemical potentials of the new and old phases. Finally, the effect of pressure and temperature variation on the thermodynamics of non-EDS clusters were studied. In Sec. \ref{sec:TranNucl} we proceed to develop the kinetics of cluster growth and decay subject to such thermodynamics.

\section{Kinetics of nucleation}\label{sec:TranNucl}

Cluster formation is a transient phenomenon with a certain lifetime which depends on size. The Szilard model explains the cluster formation as a result of a series of consecutive attachments and detachments of single monomers. It describes the kinetics of nucleation, the early stage of growth and even the Ostwald-ripening regime \cite{vetter2013} as they are mainly driven by gaining and losing monomers. The Szilard model is expressed by

\begin{align}
\label{eq:SzilardEq}
\mathrm{For}~n=1~\mathrm{:} \nonumber \\
\dfrac{dZ_1(t)}{dt}~=~&-2 f_{1}(t)Z_{1}(t)+ 2g_{2}(t)Z_{2}(t) +\sum_{n=3}^{M}g_n(t)Z_n(t) \nonumber \\
&-\sum_{n=2}^{M}f_n(t)Z_n(t)+K_1(t)-L_1(t), \nonumber \\
\nonumber \\
\mathrm{For}~ n\geq 2~\mathrm{:} \nonumber \\
\dfrac{dZ_n(t)}{dt}~=~&f_{n-1}(t)Z_{n-1}(t)-g_{n}(t)Z_{n}(t)-f_{n}(t)Z_{n}(t)  \nonumber \\
& +g_{n+1}(t)Z_{n+1}(t)+K_n(t)-L_n(t) ,
\end{align}

\noindent where $f_n(t)$ and $g_n(t)$ are attachment and detachment frequencies at time $t$, $Z_n(t)$ is the concentration of $n$-sized clusters, and $K_n(t)-L_n(t) $ reflects the non-aggregative change in the concentration of the cluster size $n$ in an open system. $K_n(t)$ is the inwards flux of $n$-sized clusters to the system from the bath and $L_n(t)$ is the outwards flux of $n$-sized clusters to the bath from the system. The Szilard model is a discrete equation. The truncated second order Taylor expansion of this discrete equation about point $n$ produces the continuous form of the Szilard model which is known as the Fokker-Planck Equation (FPE) and reads

\begin{eqnarray}
\label{eq:FPEq}
\dfrac{\partial Z(n,t)}{\partial t}~&=&-\dfrac{\partial }{\partial n}\left( v(n,t)Z(n,t)-\dfrac{1}{2}\dfrac{\partial \left[d(n,t)Z(n,t)\right]}{\partial n}\right) \nonumber \\
&&+K(n,t)-L(n,t),
\end{eqnarray}

\noindent where $v(n,t)$ and $d(n,t)$  are given by

\begin{eqnarray}
\label{eq:FPEqFreqs}
&&  v(n,t)~=~f(n,t)~-~g(n,t) \\
&&  d(n,t)~=~f(n,t)~+~g(n,t).
\end{eqnarray}

\noindent $v(n,t)$ is the drift velocity along the size axis, known as the mean growth rate, specifying the rate of deterministic incrementation of the cluster size $n$. $d(n,t)$ is the rate of random change of cluster size along the size axis (dispersion of cluster size along the size axis). The FPE is computationally favoured if the concentration of large clusters is desired. However, because of approximation in the derivation of FPE, it is inaccurate with respect to the Szilard equation at small clusters. Therefore a hybrid model is envisaged to take advantages of both discrete and continuous description of the cluster dynamics \cite{ozkan2000}. Subsequently, the cluster size axis $n$ is divided up to two sections, a discrete part $n=1, \dots, N$ and a continuous part $n=[N+1,M]$ where $N$ is the boundary between discrete and continuous sections and $M$ is the largest cluster size postulated. $N$ is chosen such that the simulation results are independent of this choice and the FPE numerically converges to the result of Szilard model. The boundary condition of the continuity of cluster flux is applied at the transition point between discrete and continuous model. The cluster flux along the size axis is defined as

\begin{eqnarray}
\label{eq:ClusterFlux}
j^c_{n}(t)~&=&~f_{n}(t)Z_{n}(t)~-~ g_{n+1}(t)Z_{n+1}(t).
\end{eqnarray}

\noindent In this study we assume that the system conserves mass. As such, for both discrete and continuous models we get $K(n,t)=L(n,t)=0 $.

 Having determined the concentration of different clusters by using the hybrid model as well as determining the nucleus size $n^*$ as the root of Eq. \ref{eq:Work11} and \ref{eq:Work11_Isotherm}, we can calculate the nucleation rate. By definition, the nucleation rate is the rate of appearance of supercritical clusters  per unit volume in the system. This is given by $J(t) = \tfrac{d\zeta(t)}{dt}$ where $\zeta(t) = \sum_{n>n^*(t)} Z_n(t)$. This is the generic definition of the nucleation rate and can be used for the non-stationary state of the old phase produced due to acoustic wave propagation.  The unknown quantities that should be determined now are the attachment and detachment frequencies.

\subsection{Monomer Attachment Frequency}
Monomer attachment to a condensed-phase cluster depends on the state of the old phase. The governing mechanism of monomer attachment is mass transfer. This usually occurs through three main mechanisms \cite{kashchiev2000}: i) direct impingement of molecules, ii) volume or surface diffusion of molecules and iii) transfer of molecules through the interface of cluster with old phase. The direct impingement is the governing mass transport mechanism when the old phase is gaseous. In the case of homogeneous nucleation in liquid or solid solutions, the main method of monomer attachment is volume diffusion. The interface-transfer method plays an important role for nucleation of clusters, solids or liquids, in a condensed old phase, e.g. melts or solutions. The diffusion mechanism may depend on the cluster size, e.g. for small clusters interface transfer may be dominant and once the cluster has grown enough the volume transfer becomes more important. If the homogeneous nucleation of solids in a dilute solution exposed to acoustic wave is the matter of concern, we postulate that volume diffusion is the main monomer attachment mechanism.

Volume diffusion can be modelled based on two different approaches: i) continuum approach, i.e. modelling the conservation of condensable mass in a supersaturated solution, and ii) atomic approach, i.e. using a random-walk model to determine the probability of collision of a monomer with a cluster and estimating the attachment frequency accordingly. Using the first approach, the attachment frequency of monomers to a spherical $n$-sized cluster in the condensed phase is given by \cite{kashchiev2000}

\begin{equation}
\label{eq:AttFreq}
f_{n}(t) = k_f(n_n) Z_1(t),
\end{equation}

\noindent where

\begin{equation}
\label{eq:KfAttachFreq}
k_f(n_n) = 4\pi \alpha_n DR_0 (1+n_n^{-\frac{1}{3}})(1+n_n^{\frac{1}{3}}),
\end{equation}

\noindent where $\alpha_n$ is the sticking coefficient which is nearly unity in a dilute solution and $D$ is the diffusivity of a monomer in the old phase. Here we assumed that both cluster and monomers are mobile and diffusing through the medium. This is implemented by using the effective diffusivity and radius for collision between a monomer and an $n$-sized cluster, as shown by Smoluchowski. The diffusivity of a cluster was estimated based on the Stokes-Einstein equation. $k_f$ resembles the collision kernel of a monomer with an $n$-mer in the Smoluchowski coagulation equation. This notion may be employed to generalise this equation for the case of non-spherical clusters by making a modification of the collision kernel using the fractal dimension of the cluster.\cite{ziff1985} These equations are valid for both discrete and continuous cluster size variable $n$.

\subsection{Monomer Detachment Frequency} \label{chap:MonDet}

The rate at which monomers detach from an $n$-sized cluster depends on the characteristics of the clusters rather than properties of the bulk new phase. This rate can be estimated following the Zeldovich method which integrates the thermodynamics under equilibrium condition into the cluster dynamics. At the thermodynamic equilibrium state, a balance between the number of monomers gained and lost by two adjacent clusters on the size axis, i.e. $j^c_{n, eq}=0$, should hold. The generalised form of the Zeldovich method for the case of time-variable supersaturation and a quasi-equilibrium condition, reads\cite{kashchiev1969}

 \begin{equation}
 \label{eq:DetachFreq_Disc}
 g_n(t)~=~f_{n-1}(t)~\exp \left( \dfrac{\Delta \Omega_n(t)-\Delta \Omega_{n-1}(t)}{k_BT(t)} \right),
 \end{equation}

  \noindent and this equation for the case of continuous cluster size $n$ becomes

 \begin{equation}
  \label{eq:DetachFreq_Cont}
  g(n,t)~=~f(n,t)~\exp \left( \dfrac{\dfrac{\partial }{\partial n}\Delta \Omega(n,t)}{k_BT(t)} \right).
  \end{equation}

\noindent For the sake of brevity, the time variable $t$ will not be noted in the following equations while all parameters are considered to be time-dependent. Substituting Eq. \ref{eq:Work9} into above equation results in

\begin{eqnarray}
\label{eq:DetachFreq_Cont1}
g(n,p,T,x)~&=&f(n,x)  \exp \left( \dfrac{1}{k_BT} \dfrac{\partial \Delta \Omega_0}{\partial n}  \right)  \nonumber \\
&& \times \exp \left(\dfrac{1}{k_BT} \int^{T_0}_T \Delta s^\prime_{exc}dT\right)  \nonumber \\
&& \times \exp \left(\dfrac{1}{k_BT} \int^{p_0}_p\Delta n^\prime_{exc} \nu dp \right).
\end{eqnarray}

\noindent The minus sign before the integrals in Eq. \ref{eq:Work9} are removed here by reversing the integration limits. This equation manifests the effect of a change in temperature and pressure on the detachment frequency of monomers from a cluster of size $n$. In our case where we are interested in investigating the effect of an acoustic wave and cavitation on nucleation and growth, this equation gives the full picture within the framework of cluster dynamics by accounting for the effect of pressure fluctuation and temperature variation due to absorption or cavitation of a bubble and mass transportation via pressure diffusion. If we use the same reference state and \emph{the new surface} as before, after some manipulations we obtain

\begin{eqnarray}
\label{eq:DetachFreq_Cont2}
g(n,p,T,x)~&=& f(n,x)~ r_0^ {\left[ \tfrac{T_0}{T } \left( h(n_n) - 1\right) \right]}  \nonumber \\
&& \times \exp \left ( \dfrac{2}{3}\dfrac{ a_0 \gamma_{\scriptscriptstyle \infty}   }{k_B T}  n_n^{-\tfrac{1}{3}}h^{'}(n_n)  \right )  \nonumber \\
&& \times  \exp \left(\dfrac{1}{k_BT} \int^{T_0}_T \Delta s^\prime_{exc}dT\right)  \nonumber \\
&& \times  \exp \left(\dfrac{1}{k_BT} \int^{p_0}_p\Delta n^\prime_{exc} \nu dp \right).
\end{eqnarray}

So far, we considered the cluster size $n$ to be a continuous variable. It is shown in Appendix \ref{sec:AppxDisctForm} that equations derived for the detachment frequency for the case of continuous $n$ can also be used for the case of discrete representation of cluster formation work. Consequently Eqs.  \ref{eq:DetachFreq_Cont1} and \ref{eq:DetachFreq_Cont2} can be employed in conjunction with the Szilard model, too.

\subsubsection{Incompressible solution and isothermal condition} \label{sec:DetachFreq_Isotherm}

The nucleation work and nucleus size in an incompressible solution which is exposed to an acoustic wave was studied in Sec. \ref{sec:IsoThermNucl}. Here we are interested in calculating the attachment and detachment frequencies under this condition. Given the volume diffusion mechanism, the diffusivity and concentration of monomers are the main factors affecting the attachment rate of monomers to an $n$-mer. The effect of pressure on diffusivity is almost negligible due to weak pressure dependence of viscosity and incompressiblity of solution. Concentration of monomers can be spatially influenced because of mass transportation due to pressure diffusion. This effect is negligible in low and medium pressure magnitudes. Nevertheless, in strong acoustic fields and specially in the vicinity of an oscillating surface, e.g. near the wall of an inertially collapsing bubble, mass transportation can be significant and should be accounted for. \cite{louisnard2007}

An acoustic wave propagating in a solution alters the thermodynamic state and consequently changes the detachment frequency, as demonstrated in Eq. \ref{eq:DetachFreq_Cont2}. In the case of an isothermal condition and pressure independent partial molecular density, this equation simplifies to

\begin{eqnarray}
\label{eq:DetachFreq_Isotherm}
g(n,p,x)~&=& f(n,x)~ r_0^ {\left( h(n_n) - 1\right) } \exp \left ( \dfrac{2}{3}\dfrac{ a_0 \gamma_{\scriptscriptstyle \infty}   }{k_B T}  n_n^{-\tfrac{1}{3}}h^{'}(n_n)  \right )   \nonumber \\
&& \times \exp \left( - \dfrac{\nu \Delta p}{k_BT} \Delta n^\prime_{exc} \right).
\end{eqnarray}

\noindent Subsequently, approximating molar concentration with the concentration of monomers, i.e. $r_0=\frac{x_0}{x_e}=\frac{Z_1}{C_e}$ where $C_e$ is the solubility at the reference state, we arrive at

\begin{eqnarray}
\label{eq:DetachFreq_Isotherm2}
g(n,p,x) ~&=& k_f(n_n)~ C_e r_0^ { h(n_n) } \exp \left ( \dfrac{2}{3}\dfrac{ a_0 \gamma_{\scriptscriptstyle \infty}   }{k_B T}  n_n^{-\tfrac{1}{3}}h^{'}(n_n)  \right )   \nonumber \\
&& \times \exp \left( - \dfrac{\nu \Delta p}{k_BT} \Delta n^\prime_{exc} \right),
\end{eqnarray}

\noindent and the latest assumption is justified since the concentration of monomers in the system at the initial time is significantly greater than that of $n$-mers.

\section{Results and discussion}\label{sec:ResultDisc}
We have established all the required equations to determine the kinetics of nucleation while accounting for the effect of fluctuations in the thermodynamic state of the old phase. In the first part of this section we examine the new development by applying it to the test case of water droplet nucleation from the gas phase. The model of water was chosen given the fact that homogeneous nucleation of vapour is very well studied both experimentally and theoretically. Subsequently, having validated the model and numerical implementation, we will evaluate the effect of an acoustic wave on crystallisation in an aqueous solution in Sec. \ref{sec:CrystallisationResults}. We will use these results to explain some experimental trends reported in the literature. Since the majority of experimental works do not define all the necessary parameters of both the acoustic field and crystallisation, a direct comparison with the sonocrystallisation data seems impractical. In addition, acoustic cavitation usually happens prior to or concurrent with crystallisation which is often not characterised in experiments.

For the numerical computations, the FPE (Eq. \ref{eq:FPEq}) is discretized and solved together with the discrete Szilard equation (Eq. \ref{eq:SzilardEq}) using a variable ODE solver. The details of numerical implementation are expressed elsewhere.\cite{chang1970, vetter2013, brown1989} We define the dimensionless time variable $\tau$ here as follows: $\tau=k_t t$ where $k_t=4\pi R_0 D C_e$ in the case of attachments governed by the volume diffusion process. Regarding the initial condition, following our previous discussion we assume only monomers are present in the system and bath initially. The presence of $n$-sized clusters, $2\leq n \leq 4$, in the initial condition may change the nucleation rate by less than one order of magnitude.\cite{kozisek2005} Consequently, the considered initial condition is adequately reasonable for our work. For all the simulations, we consider supersaturation is time-varying and the system is closed.

\subsection{Water droplet nucleation} \label{sec:WaterDropletResults}

The excess free energies of water droplets of different sizes have recently been calculated by means of a statistical mechanical approach at a temperature $T=300~\mathrm{K}$.\cite{samsonov2003,lau2015} Considering that calculations using the TIP4P/2005 molecular model could successfully estimate the surface energy in agreement with experiments,\cite{wilhelmsen2015} we use the calculations of Lau et al. \cite{lau2015} to validate our model following the ensuing procedure: i) we deduce the values of $\lambda$ by comparing the effective surface tension determined by our model, Eq. \ref{eq:GammaEff}, with those obtained from statistical mechanics, and ii) use these results to calculate nucleation rates and compare them against experimental results at $T=300~ \mathrm{K}$ obtained by Brus et al.. \cite{brus2008,brus2009}

\begin{figure}[ht!]
\centering
\includegraphics{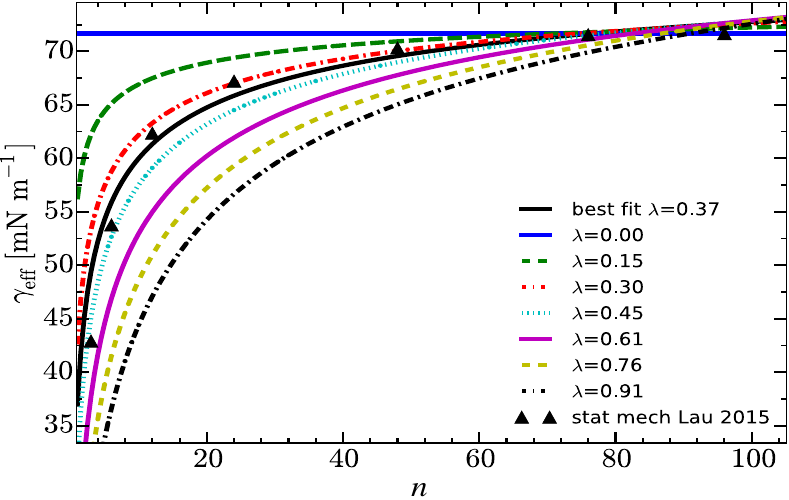}
\caption{ Effective surface energy ($\gamma_{\mathrm{\scriptscriptstyle eff}}$) at different $\lambda$ calculated by Eq. \ref{eq:GammaEff} at $T=300~ \mathrm{K}$. $\filledmedtriangleup$: Statistical mechanical simulations\cite{lau2015} at cluster size of $n=3,6,12,24,48,76,96$. Solid black curve shows the best fit to statistical mechanical simulations.}
\label{fig:gammaeff}
\end{figure}

Figure \ref{fig:gammaeff} shows the effective surface tension at different values of $\lambda$ calculated at experimental supersaturation $S=3.77$ at $T=300~ \mathrm{K}$. We can see that at the very small clusters of size $n=3 ~\mathrm{and} ~6$, the best fit is achieved at $\lambda=0.45$ whereas for larger clusters ($n>12$) the curve with $\lambda=0.37$ happens to give the best agreement with statistical mechanic results. This may suggest that $\lambda$ is size-dependent analogous to the Tolman length \cite{talanquer1995,lu2005,granasy1998} for the surface of tension as the dividing surface. Nevertheless, since with $\lambda=0.37$ we achieve acceptable approximation of size-dependent surface energy with respect to statistical mechanical simulations over a wide range of cluster size, we take this value for our calculations at this condition. The efficacy of this choice will be evaluated by comparing the calculated kinetics of nucleation with the experiments.

The same procedure was repeated for all experimental supersaturations at $T=300~ \mathrm{K}$ reported by Brus et al. \cite{brus2008,brus2009} and we observed the same trend showing that $\lambda$ is larger for small clusters ($n<12$) and decreases for larger clusters ($n>12$). Likewise, in the case of the surface of tension as the dividing surface, the similar tendency of size-dependence of Tolman length, i.e. inverse relationship with droplet size, and supersaturation dependence at a constant temperature were also reported. \cite{lu2005,granasy1998}

Having determined the parameter $\lambda$ for a range of supersaturations at temperature $T=300~ \mathrm{K}$, we can now calculate the kinetics of water droplet nucleation by solving the hybrid model. Considering that for the gaseous old phase, clusters are much smaller than the mean free path in the gas phase, the attachment rate should correspond to a particle flux modelled by the gas kinetic theory. Consequently, the attachment frequency reads \cite{kashchiev2000, vehkamaki2006}

\begin{eqnarray}
\label{eq:GKTAttFreq}
 f_{n}(t) &=& \alpha_n a_0 \sqrt{ \dfrac{k_B T}{2 \pi m_0} } (1 + \dfrac{1}{n})^{\frac{1}{2}}   (1+n_n^{\frac{1}{3}})^2 Z_1(t) \nonumber \\
& = & k_{f,g} (n_n) Z_1(t)
\end{eqnarray}

\noindent where $k_{f,g}(n_n) =  \alpha_n a_0 \sqrt{ \tfrac{k_B T}{2 \pi m_0} } (1 + \tfrac{1}{n})^{\tfrac{1}{2}}   (1+n_n^{\frac{1}{3}})^2$. Based on the definition of a Gibbs surface, the mass of a cluster of size $n$ is given by $m_n=n m_0$, where $m_0$ is the mass of a monomer in the new phase, but the radius of cluster is defined by $n_n$. Subsequently, if we use $k_{f,g}(n_n)$  instead of $k_{f}(n_n)$ given in Eq. \ref{eq:KfAttachFreq}, all the previous equations are applicable and can be used for this exercise. The time non-dimensionalisation coefficient also changes to $k_t=a_0   \sqrt{ \tfrac{k_B T}{2\pi m_0}} \tfrac{p_{ve}}{k_B T}$ where $p_{ve}$ is the equilibrium vapour pressure. At $T=300~ \mathrm{K}$, we have $k_t=5.9\times10^7~\mathrm{s^{-1}}$.

\begin{figure}[ht!]
\centering
\includegraphics{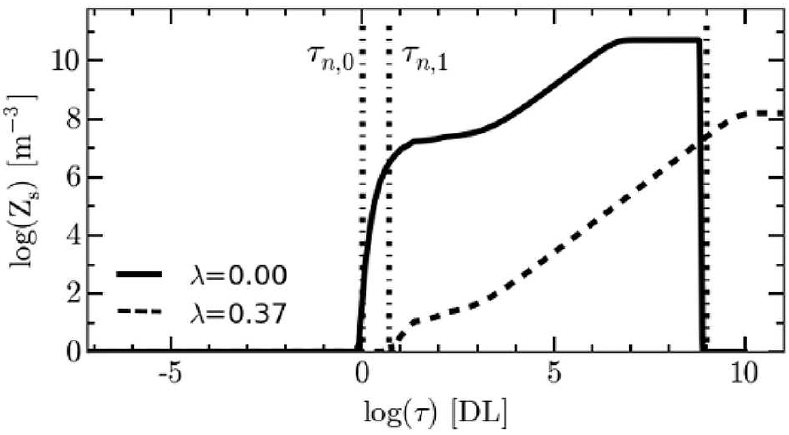}
\caption{ Logarithmic concentration (logarithm in base 10 in all figures) of supercritical clusters at two different $\lambda$ values over time. Vertical lines labelled $\tau_{\scriptscriptstyle n,0}$ and $\tau_{\scriptscriptstyle n,1}$ indicate the beginning of the nucleation stage in models with $\lambda=0$ and $\lambda=0.37$, respectively. The unlabelled vertical line indicates the end of nucleation and monomer-driven growth of supercritical clusters in the case of $\lambda=0$ while nucleation is still ongoing in the case of $\lambda=0.37$. This is due to a faster nucleation rate for $\lambda=0$ which leads to quicker depletion of the imposed supersaturation of monomers. }
\label{fig:Zsupnucovertime}
\end{figure}

\begin{figure}[h!]
\centering
\includegraphics{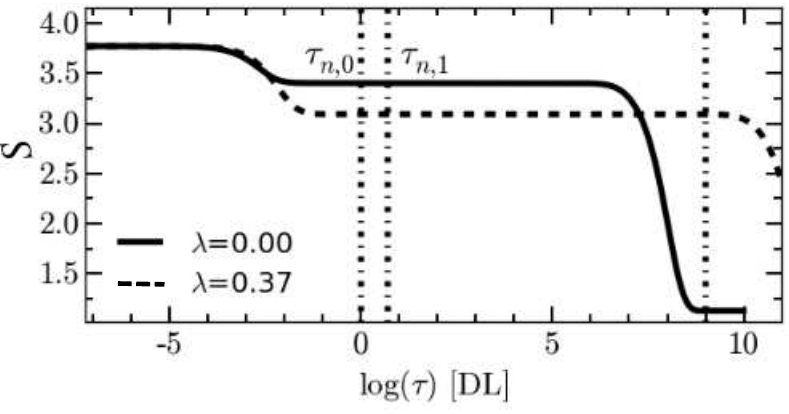}
\caption{ Supersaturation at two different $\lambda$ values over time. See caption of Figure \ref{fig:Zsupnucovertime} for details.  }
\label{fig:Sovertime}
\end{figure}

The physicochemical properties of a vapour mixture are determined by equations provided in Table 1 of Brus et al.. \cite{brus2008} The hybrid model is then numerically solved and the nucleation rate  is calculated. Figures \ref{fig:Zsupnucovertime} and \ref{fig:Sovertime} depict variation in concentration of supercritical clusters ($\zeta(\tau)$) and supersaturation over time, respectively, for two cases where clusters are defined by the EDS ($\lambda=0$ ) and the surface with $\lambda=0.37$. The equilibrium monomer concentration of $C_e=p_{ve}/k_BT$ is used to determine the concentrations $Z(n)$ in the supersaturated state. The nucleation time is identified by the appearance of the first ten supercritical clusters, i.e. $\mathrm{log}(Z_{s})\geq1$, which happens around $\tau_{\scriptscriptstyle n,0}=1$ and $\tau_{\scriptscriptstyle n,1}=7$ for these EDS and non-EDS cases, respectively, and are indicated by vertical dashed-dotted lines, see Figure \ref{fig:Zsupnucovertime}. The system is closed and therefore the total mass is a constant. As a result, the condensation depletes the monomer supersaturation and terminates the nucleation and monomer-driven growth stages around $\tau=10^9$ for the classical cluster case. At this moment the concentration of monomers drops drastically ($S\approx1$) whereas it lasts longer for the non-EDS clusters. This implies that this new model with  $\lambda=0.37$ predicts lower nucleation rate than the CNT. The sharp fall marked by the vertical line at the right hand side is due to the way we define $\zeta(\tau)$, see above, and does not hold physically. Since supersaturation drops to almost unity, the critical cluster size mathematically tends to infinity and therefore all the previously made clusters become subcritical which brings about this abrupt drop of $\zeta(\tau)$.

\begin{figure}[h!]
\centering
\includegraphics{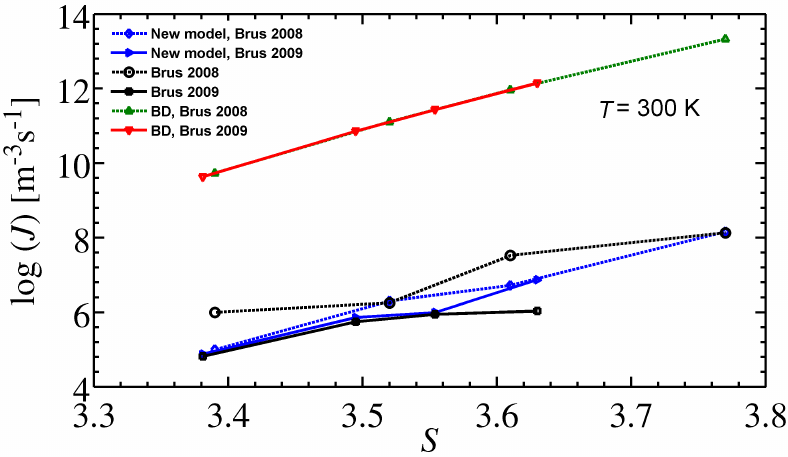}
\caption{Logarithm of nucleation rate versus supersaturation at $T=300~ \mathrm{K}$. Nucleation rate calculated by our new model using $\lambda$ values determined from statistical mechanical calculation of Gabriel et. al.\cite{lau2015} at the data points of Brus 2008 and 2009 (solid line with $\filleddiamond$ and  dashed line with $\filledmedtriangleright$, respectively). The experimental results of Brus et. al. 2008 and 2009\cite{brus2008,brus2009} are also shown (solid line with $\bullet$ and dashed line with $\filledmedsquare$, respectively). Nucleation rate determined by BD model at the data points of Brus 2008 and 2009 ($\filledmedtriangleup$ and $ \filledmedtriangledown$, respectively).  }
\label{fig:JSimExpBD}
\end{figure}

Repeating these calculations for all experimental supersaturations, we determine the stationary nucleation rates for all these conditions. These results together with the experimental nucleation rate and values obtained by Becker-D\"oring (BD) model are plotted in Figure \ref{fig:JSimExpBD}. The main difference between our model and the BD is due to the non-EDS definition of clusters in our model which allows a more accurate estimation of the excess free energy whereas the BD model calculates the stationary nucleation rate using a classical definition of a cluster. This also leads to a different critical cluster size and consequently a different nucleation rate. Additionally, the hybrid Szilard and FPE model determine the kinetics of nucleation for a stationary as well as a time variable non-stationary system. This is important because in practice the supersaturation imposed on the system is always time variable. We used the same physicochemical properties of the vapour mixture and experimental supersaturations in the BD.

\begin{figure}[h!]
\centering
\includegraphics{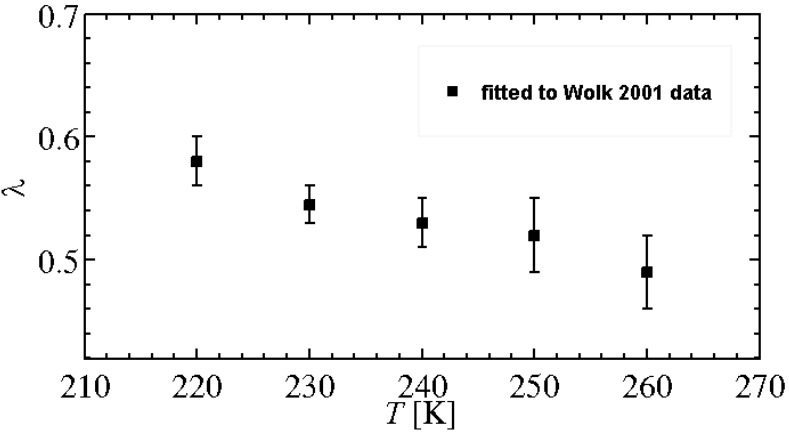}
\caption{ $\lambda$ calculated from the experimental nucleation rate\cite{wolk2001} for water nucleation at different temperatures. The error bars show the range in $\lambda$ at a specific temperature as a function of supersaturation. The lower and upper limits correspond to the smallest and largest experimental supersaturations at a specific temperature, respectively.    }
\label{fig:lambfitted}
\end{figure}

The agreement between predicted nucleation rate and experimental values are very good. Though the dividing surface we used to define clusters has the property of size-independent surface tension, the effective surface tension of this surface is size, temperature and supersaturation dependent, see Eq. \ref{eq:GammaEff}. This is attributed to the fact that in our model, a cluster of size $n$ can take on different combinations of $n_n$ and $n_\sigma$ due to the arbitrary placement of dividing surface contrary to a cluster defined by EDS or surface of tension. Therefore we are able to reproduce $\gamma_{\mathrm{\scriptscriptstyle eff}}$ by choosing the location of the dividing surface appropriately, see Figure \ref{fig:gammaeff}. It seems that this important characteristic corrects some of the shortcomings of CNT, at least for water at $T=300~ \mathrm{K}$. We should not, however, dismiss the chance of a coincidental close agreement between our numerical results and experimental values since this model does not account for non-idealities in the gas phase and compressibility of the liquid phase. The lack of molecular simulation results at other temperatures does not allow us at this stage to extend these calculations to lower temperatures. Nevertheless, we deduced the values of $\lambda$ from experimental data of W\"olk et al.,\cite{wolk2001} shown in Figure \ref{fig:lambfitted}. We can see the gradual descent of $\lambda$ with temperature rise. A similar trend was observed and reported for the Tolman length too. \cite{talanquer1995, holten2005}

We demonstrated in this section that using a specific non-EDS dividing surface together with a Gibbs droplet model in a general format can better predict the kinetics of nucleation compared to the CNT. This supports our intention to study the effect of pressure fluctuations on the nucleation process by means of a more generic model than the CNT. These results are elaborated in Sec. \ref{sec:CrystallisationResults}.

\subsection{Effect of pressure fluctuation on crystallisation} \label{sec:CrystallisationResults}

The effect of pressure fluctuation on the clustering work and the kinetics of nucleation in the system are determined by Eqs. \ref{eq:WorkdiffCondPress} and \ref{eq:DetachFreq_Cont2}, respectively. Employing these equations under isothermal condition in our hybrid cluster dynamics model, we can study nucleation, the early stage of growth and also the Ostwald-ripening phenomenon in a system undergoing pressure variation.

We consider a single closed system in the bath with a time varying pressure. The local pressure can then be written as $p = p_{\scriptscriptstyle a}~+~p_{\scriptscriptstyle 0}$ where $p_{ \scriptscriptstyle 0}$ is the ambient pressure at the reference state and $p_{\scriptscriptstyle a} = p_{\scriptscriptstyle m} \cos(2\pi f t)$ is the acoustic pressure in the system with magnitude $p_{\scriptscriptstyle m}$ and frequency $f$.

Here we report simulations performed using different sets of parameters. Acoustic pressure and frequency are varied from $1~\mathrm{MPa}$ to $50~\mathrm{MPa}$ and $20~\mathrm{kHz}$ to $2~\mathrm{MHz}$, respectively. This range of acoustic parameters pertains to experimental amplitude and frequency of ultrasound waves generated by different ultrasonic transducers, e.g. planar and high intensity focused, in sonocrystallisation experiments. Further, we studied the effect of the parameter $\lambda$ on the kinetics of nucleation under isothermal pressure perturbation too. With regard to the solution properties,  we use the generic physicochemical properties of a sparingly soluble salt in an aqueous solution at room temperature ($T = 293~\mathrm{K}$) provided in Table 6.1 of Reference \onlinecite {kashchiev2000}. Following this reference, we consider the new phase to be denser than the old phase with a typical value of $\Delta \nu = 10^{-28}~\mathrm{m^{3}}$. This gives $k_\rho>0$. Furthermore, we have the time non-dimensionalisation constant of $k_t=2.9\times10^5~\mathrm{s^{-1}}$. Unless otherwise stated, all the following simulations are conducted with $\lambda=0.35$ which is an average value of $\lambda$ formerly obtained for water droplet formation at $T = 300~\mathrm{K}$. This choice is improvised assuming the surface energy of clusters in a dilute aqueous solution shows a similar size dependence at the same temperature as water droplets. Nevertheless, we investigate the effect of different values of $\lambda$ in Sec. \ref{sec:SimLambdaEff}.

\subsubsection{Simulations at varying pressure magnitude and frequency }
\begin{figure}[ht]
\centering
\includegraphics{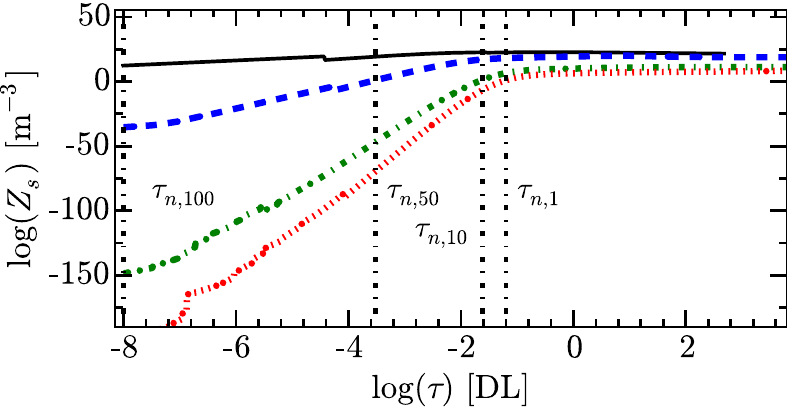}
\caption{ Concentration of supercritical clusters at different pressure magnitudes over time with $\lambda=0.35$. Vertical lines labelled $\tau_{\scriptscriptstyle n,100}$ to $\tau_{\scriptscriptstyle n,1}$ indicate the beginning of the nucleation stage for different static pressures of old phase. Static pressure decreases from the black curve ($100~\mathrm{MPa}$) at the top to the red curve at the bottom ($1~\mathrm{MPa}$).}
\label{fig:Zsupnucovertime_lam35}
\end{figure}

Initially, we investigate the effect of the magnitude of static pressure on crystallisation. This is obtained by setting $f=0$ and keeping $p_{\scriptscriptstyle 0}$ constant. Figure \ref{fig:Zsupnucovertime_lam35} shows changes in concentration of supercritical clusters over time at different pressure magnitudes of  $1~\mathrm{MPa}, 10~\mathrm{MPa}, 50~\mathrm{MPa}$ and $100~\mathrm{MPa}$. The dashed vertical lines illustrate the nucleation time lag at different pressure magnitudes. We can see that in the case of positive $k_\rho$, $\tau_n$ has an inverse relation with pressure magnitude.  For example, the nucleation time lag reduces by more than six orders of magnitude as the pressure magnitude increases only by one order from $10~\mathrm{MPa}$ to $100~\mathrm{MPa}$ ($\tau_{\scriptscriptstyle n,100}\approx 10^{-8}$ which gives  $t_{\scriptscriptstyle n,100}\approx 30~ \mathrm{fs}$). A similar trend between the pressure magnitude $p_m$ and the experimental lifetime of superheated xenon, oxygen and argon liquids was reported in the literature too. \cite{baidakov1981,baidakov2009}

\begin{figure}[ht]
	\centering
	\includegraphics{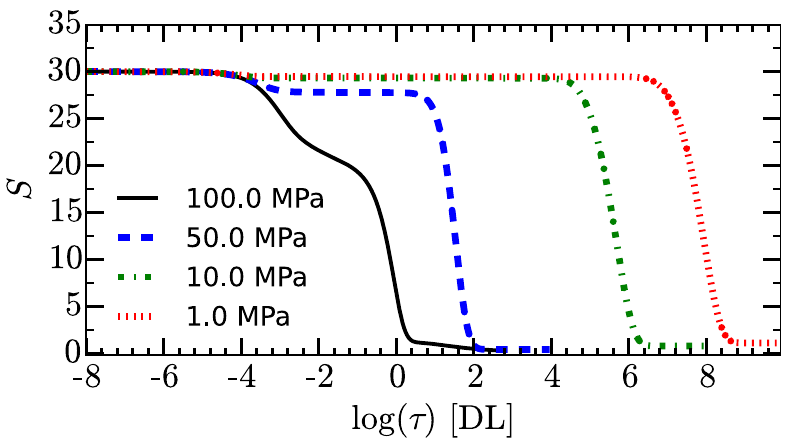}
	\caption{ Supersaturation at different pressure magnitudes over time with $\lambda=0.35$. }
	\label{fig:Sovertime_lam35}
\end{figure}

\begin{figure}[ht]
\centering
\includegraphics{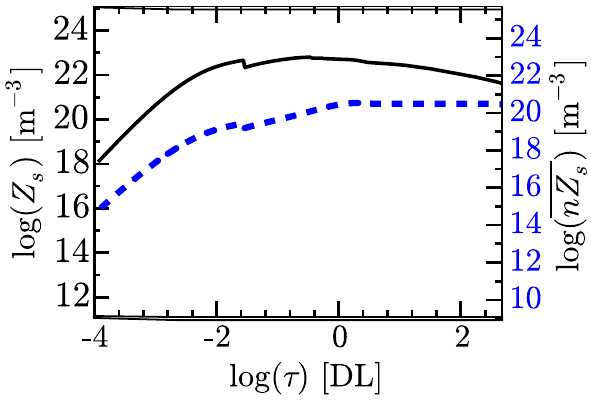}
\caption{ Concentration of supercritical clusters (solid line) and mean size of supercritical clusters (dashed line) at static pressure of $100~\mathrm{MPa}$ and with $\lambda=0.35$. Around $\mathrm{log(\tau)}=-0.7$ the concentration of supercritical clusters becomes a maximum and starts to decline whereas the mean size of supercritical clusters increases and plateaus shortly after. }
\label{fig:Zsn_ZsnAVG_Pac100MPa}
\end{figure}

\begin{figure}[ht]
	\centering
	\includegraphics{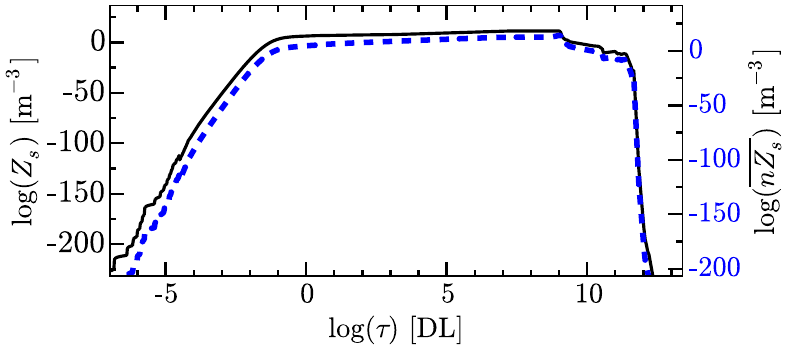}
	\caption{ Concentration of supercritical clusters (solid line) and mean size of supercritical clusters (dashed line) at static pressure of $1~\mathrm{MPa}$ and with $\lambda=0.35$. Around $\mathrm{log(\tau)}=8.8$ the supersaturation approaches unity, the concentration of supercritical clusters becomes a maximum and starts to decline. The mean size of supercritical clusters drops too but a ripening process could not be identified.}
	\label{fig:Zsn_ZsnAVG_Pac1MPa}
\end{figure}

The change in the supersaturation over time is depicted in Figure \ref{fig:Sovertime_lam35}. An increase in pressure magnitude amplifies the depletion rate of imposed monomers in a closed system. This fast nucleation rate leads to smaller supercritical clusters on average. At the highest pressure, the Ostwald ripening regime starts at roughly $\mathrm{log(\tau)}=-0.7$ where the concentration of supercritical clusters reaches its maximum and declines afterwards. We can see that the average size of supercritical clusters, however, increases after this instance which is due to the absorption  of depleted monomers from smaller clusters by larger clusters, see Figure \ref{fig:Zsn_ZsnAVG_Pac100MPa}. The Ostwald ripening, however, could not be observed when $p_{\scriptscriptstyle m} = 1~\mathrm{MPa}$, see Figure \ref{fig:Zsn_ZsnAVG_Pac1MPa}. In this case the concentration and the average size of supercritical clusters increase and sharply drop together. As a result, we expect to see a cluster size distribution (CSD) with a smaller mean value and a broader distribution in a higher magnitude excitation than a lower one.

\begin{figure}[ht!]
	\centering
	\includegraphics{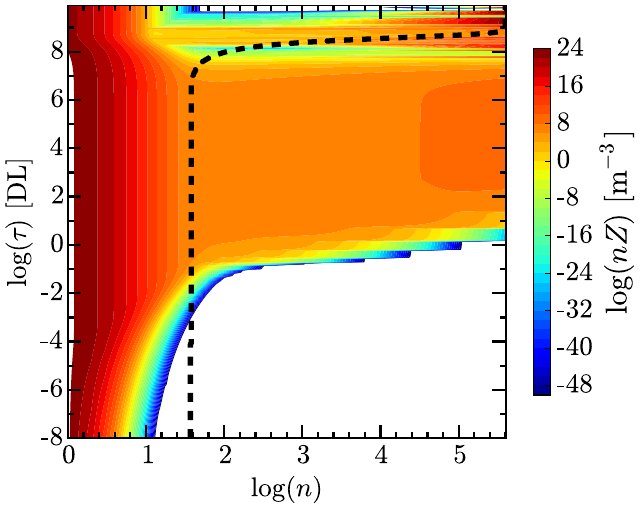}
	\caption{  Logarithmic CSD, i.e. the contour plot of logarithmic size-weighted cluster size distribution ($\mathrm{log(}nZ\mathrm{)}$), over time at the static pressure $1~\mathrm{MPa}$ with $\lambda=0.35$. The black dashed line shows the time variable size of the critical cluster. }
	\label{fig:Zsn_tau_n_Pac1MPa}
\end{figure}

\begin{figure}[ht!]
	\centering
	\includegraphics{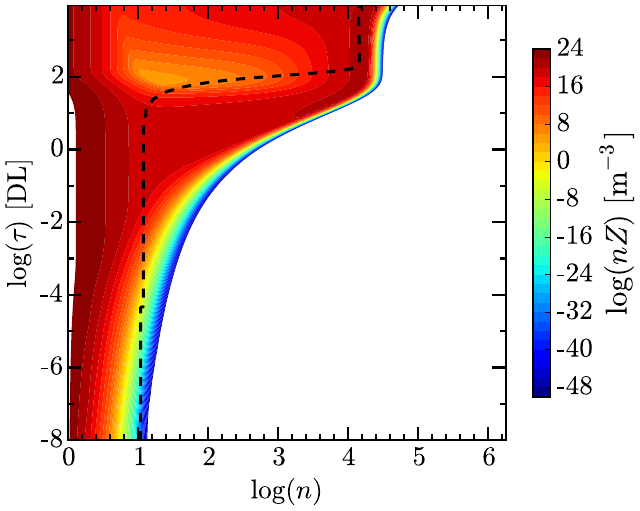}
	\caption{ Logarithmic CSD  over time at the static pressure of $50~\mathrm{MPa}$ with $\lambda=0.35$. The black dashed line shows the time variable size of the critical cluster. }
	\label{fig:Zsn_tau_n_Pac50MPa}
\end{figure}

\begin{figure}[ht!]
	\centering
	\includegraphics{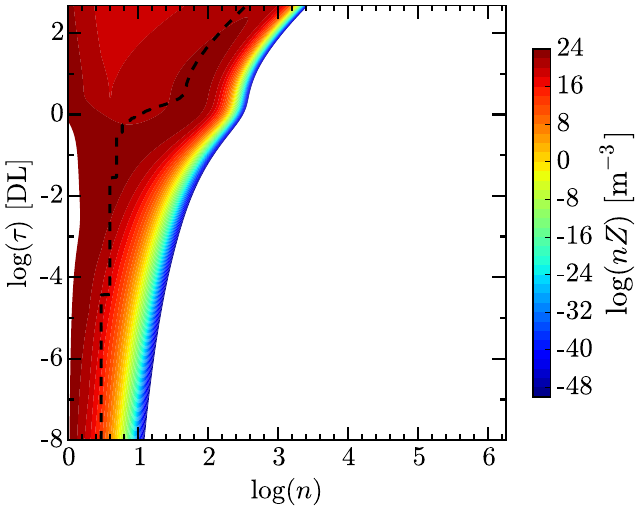}
	\caption{  Logarithmic CSD  over time at the static pressure of $100~\mathrm{MPa}$ with $\lambda=0.35$. The black dashed line shows the time variable size of the critical cluster. }
	\label{fig:Zsn_tau_n_Pac100MPa}
\end{figure}

\begin{figure}[ht!]
	\centering
	\includegraphics{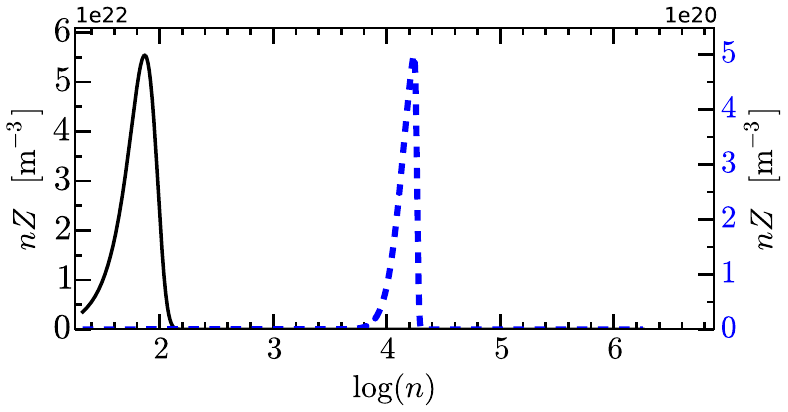}
	\caption{  The CSD at the end of nucleation stage ($S\approx1$) at two static pressures of $100~\mathrm{MPa}$, the left vertical axis, and $50~\mathrm{MPa}$, the right vertical axis. Refer to the text for details. }
	\label{fig:Z_n_Pac100_50MPa}
\end{figure}

The contour plots in Figures \ref{fig:Zsn_tau_n_Pac1MPa}-\ref{fig:Zsn_tau_n_Pac100MPa} show the size-weighted cluster size distribution at three different static pressures. The time-variable size of critical cluster is overlaid on each plot. We can obviously see that the size of critical clusters follow the same trend as supersaturation over time. The initial critical cluster sizes are $40, 16$ and $4$ at pressure magnitudes of $1~\mathrm{MPa}, ~50~\mathrm{MPa}$ and $100~\mathrm{MPa}$, respectively. Furthermore, these plots illustrate that the size of critical and mean size of supercritical clusters inversely correlate with pressure magnitude (when $k_\rho>0$). Reading the concentration of clusters at the end of the nucleation period, i.e. $S\approx1$, from this contour, we obtain the CSD under these conditions, depicted in Figure \ref{fig:Z_n_Pac100_50MPa}. This figure shows that the mean of the CSD becomes smaller as pressure increases. This is attributed to a short nucleation period due to a fast nucleation rate which causes a significant reduction in the time difference between the birth time of different stable supercritical clusters. Furthermore, the distribution becomes broader at the higher pressure magnitude which is due to enhancement of the ripening process with pressure rise (when $k_\rho>0$).

As we have seen so far, a relatively high magnitude of pressure is required to influence the nucleation process. Static pressure can be manipulated within this range and even higher experimentally using a high pressure chamber. In terms of the ultrasonic pressure oscillation at such magnitudes, a focused transducer is required as available flat transducers are unable to generate such a strong pressure field. A high intensity focused ultrasound transducer operating at high driving frequencies, e.g. $1-3~\mathrm{MHz}$, can generate high magnitude pressure oscillation at focus in water. \cite{crum2008} Such a strong high frequency acoustic wave, however, becomes distorted and turns into shock due to nonlinearities of the transducer and the wave medium. Nevertheless, our main objective in this work is to develop and study a theoretical approach for such applications and therefore we will approximate pressure oscillation by a sinusoidal wave in the following simulations. Furthermore, we only account for the direct acoustic field and exclude the emitted pressure from the potential acoustic cavitation which may occur at a setting of the acoustic field. Nevertheless, incorporating the bubble dynamics into our model, we can estimate the thermodynamics and kinetics of nucleation stimulated by the bubble dynamics too.

\begin{figure}[ht!]
\centering
\includegraphics{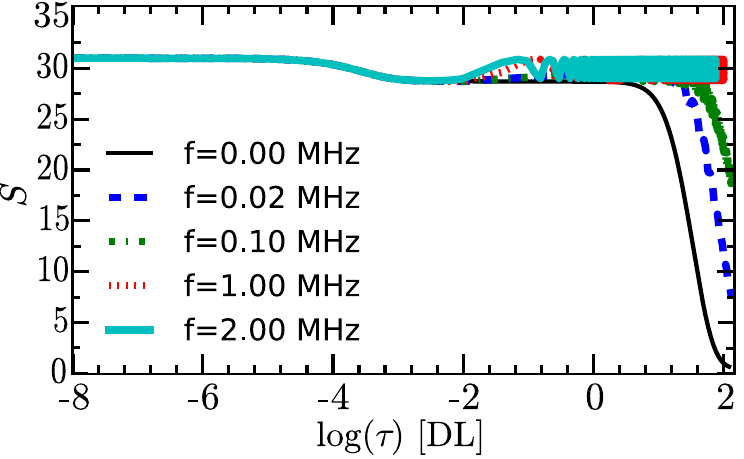}
\caption{ Supersaturation over time at different excitation frequencies and pressure magnitude of $p_{\scriptscriptstyle m}=50~\mathrm{MPa}$  with $\lambda=0.35$. }
\label{fig:Sovertime_lam35_Allfreq}
\end{figure}

\begin{figure}[ht!]
\centering
\includegraphics[width=8cm]{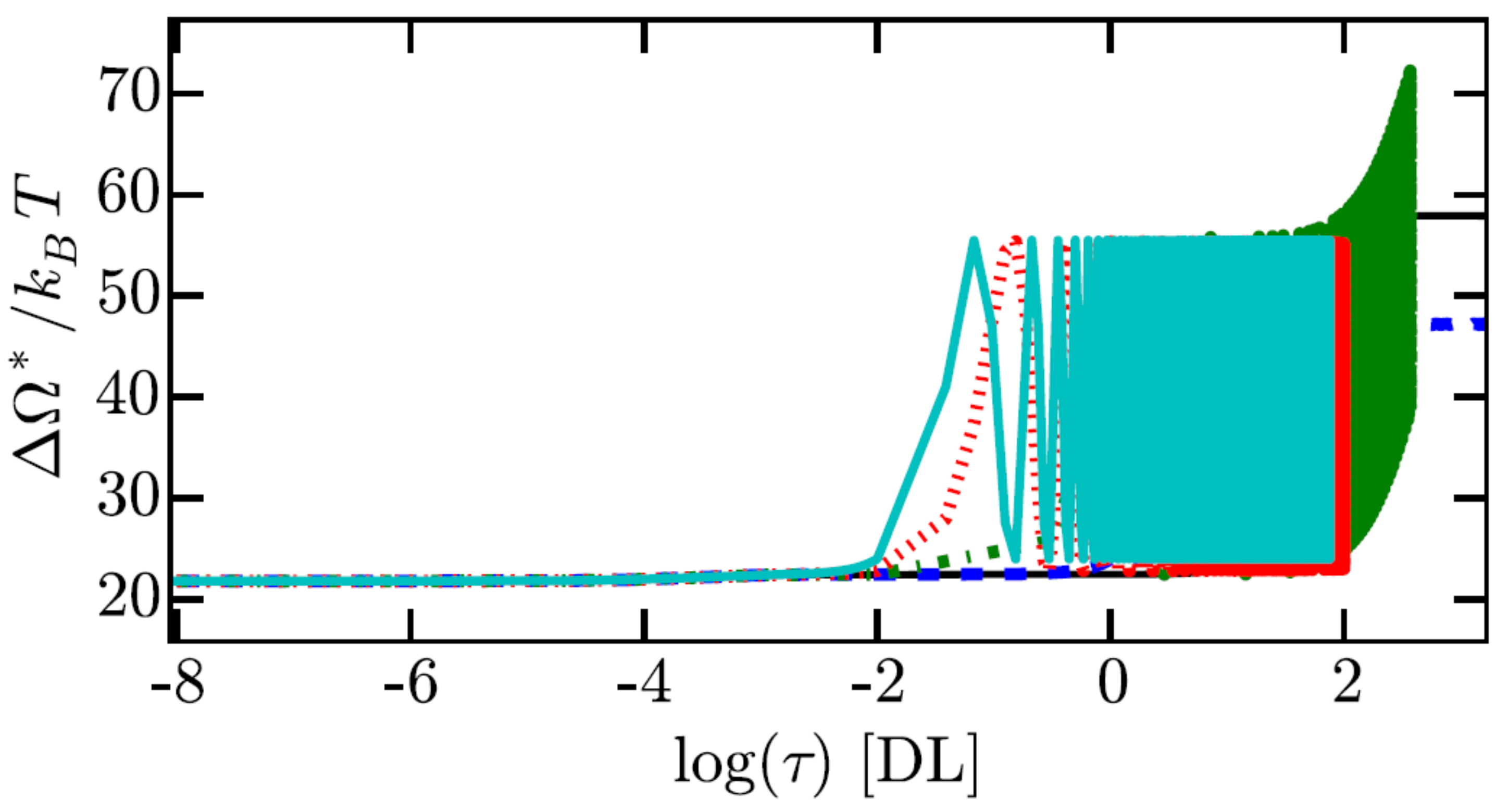}
\caption{ Nucleation work over time at different excitation frequencies and pressure magnitude of $p_{\scriptscriptstyle m}=50~\mathrm{MPa}$  with $\lambda=0.35$. The legend is the same as that of Figure \ref{fig:Sovertime_lam35_Allfreq}.}
\label{fig:DelOmegastarovertime_lam35_Allfreq}
\end{figure}

\begin{figure}[ht!]
\centering
\subfigure[]{
\centering
\includegraphics{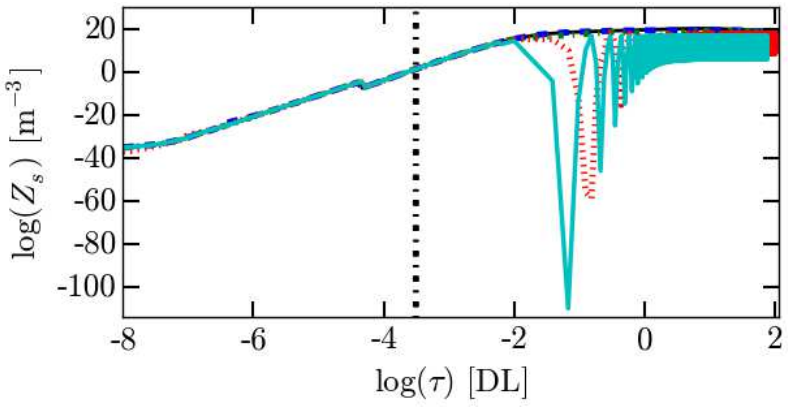}
}
\subfigure[]{
\centering
\includegraphics{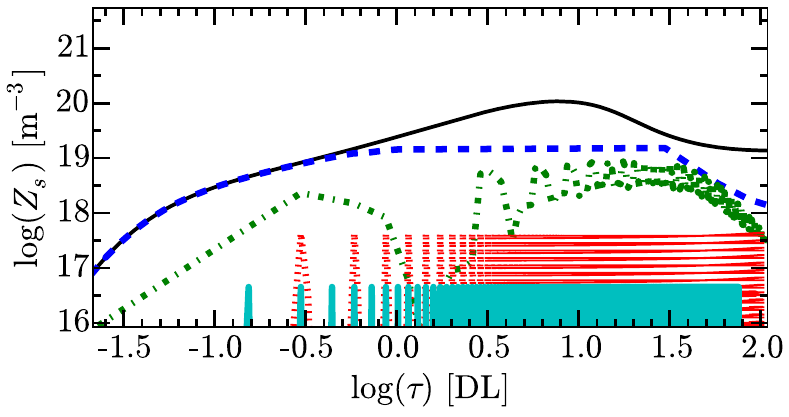}
}
\caption{ (a) Concentration of supercritical clusters over time where the old phase is exposed to  pressure fluctuation with $p_m=50~\mathrm{MPa}$ at different frequencies. The legend is the same as that of Figure \ref{fig:Sovertime_lam35_Allfreq}. (b) magnified towards the end of simulation with $f=0~\mathrm{MHz}$ (when $S\approxeq 1$). We can see that at  $f=0, ~20,~100~\mathrm{kHz}$, $Z_s$ reaches its maximum value at $\mathrm{log(}\tau\mathrm{)}=0.9, ~1.4,~1.5$, respectively. The Ostwald ripening process follows. In contrast, at higher frequencies the nucleation stage is still ongoing at this time.}
\label{fig:Zsn_tau_n_Pac50MPa_Allfreq_lam35}
\end{figure}

The simulation results of nucleation in the same aqueous solution exposed to an acoustic wave with $p_{\scriptscriptstyle m} = 50~\mathrm{MPa}$ and frequencies of $0.02, ~0.1,~ 1$ and $2~\mathrm{MHz}$ are presented below. Comparing to the static pressure condition, pressure oscillation leads to a smaller effective pressure magnitude which lowers the effective nucleation rate. This point is observed in Figure \ref{fig:Sovertime_lam35_Allfreq} where the nucleation stage ends at $\mathrm{log(\tau)}=0.8, 1.4, 1.8$ at driving frequencies of $0, 0.02$ and $0.1~\mathrm{MHz}$, respectively, whereas it is still ongoing in higher frequency oscillations. The main reason for this behaviour is the variation in the nucleation work due to pressure oscillations, see Figure \ref{fig:DelOmegastarovertime_lam35_Allfreq}, and subsequently the detachment frequency. Equation \ref{eq:WorkdiffCondPress} shows that in an isothermal process, pressure can impede or facilitate nucleation depending on the sign of $\Delta n_{exc} \Delta p$. When $\Delta n_{exc}$ is positive (i.e. the formation of a condensed phase), an isothermal increase in reference pressure reduces the nucleation work and consequently the depletion rate, Eq. \ref{eq:DetachFreq_Isotherm2}, which gives a higher nucleation rate and vice versa. This also influences the concentration of supercritical clusters, shown in Figure \ref{fig:Zsn_tau_n_Pac50MPa_Allfreq_lam35}, such that $Z_{s}$ reduces as frequency increases.

\begin{figure}[h!]
\centering
\includegraphics{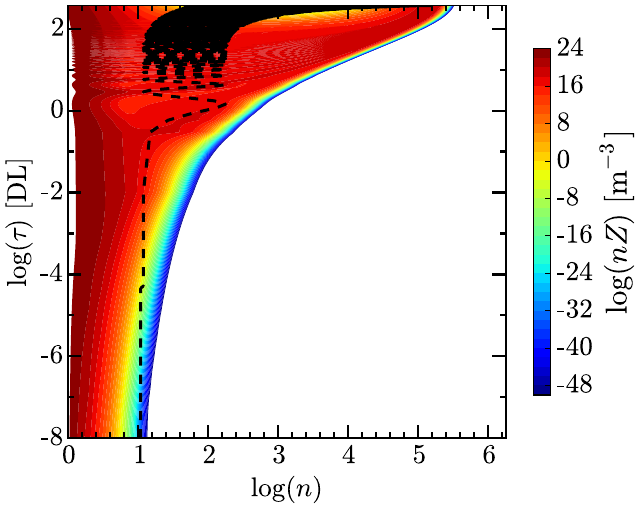}
\caption{ Logarithmic CSD over time where solution is exposed to an acoustic wave with excitation parameters of $f=100~\mathrm{kHz}$ and $p_{\scriptscriptstyle m}=50~\mathrm{MPa}$ with $\lambda=0.35$. }
\label{fig:Zsn_tau_n_Pac50MPa_100kHz}
\end{figure}

\begin{figure}[h!]
\centering
\includegraphics{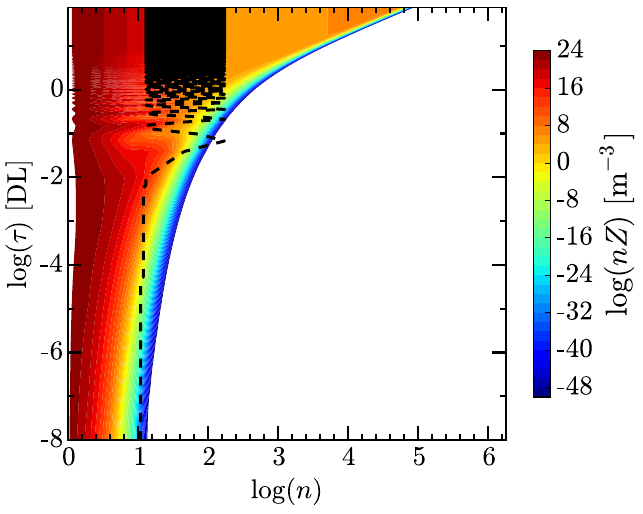}
\caption{ Logarithmic CSD over time where solution is exposed to an acoustic wave with excitation parameters of $f=2~\mathrm{MHz}$ and $p_{\scriptscriptstyle m}=50~\mathrm{MPa}$ with $\lambda=0.35$. }
\label{fig:Zsn_tau_n_Pac50MPa_2MHz}
\end{figure}

The CSD contour plots for two frequencies are shown in Figures \ref{fig:Zsn_tau_n_Pac50MPa_100kHz} and \ref{fig:Zsn_tau_n_Pac50MPa_2MHz}. Comparing them with the CSD at static pressure of $p_{\scriptscriptstyle m} = 50~\mathrm{MPa}$, we observe that supercritical clusters become more numerous, i.e. we have nonzero concentrations at $\mathrm{log(}n\mathrm{)}>4.8$. As we discussed above, their concentration, however, is reduced due to pressure oscillations. Overall, an acoustic wave causes reduction in the magnitude of the CSD at the end of nucleation and moves the mean of the CSD to a larger $n$ as frequency goes up.

\subsubsection{Simulations at different $\lambda$} \label{sec:SimLambdaEff}

\begin{figure}[ht]
\centering
\includegraphics{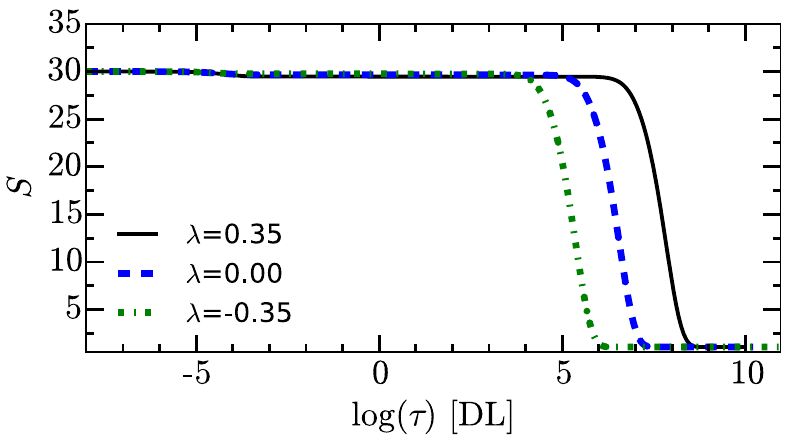}
\caption{ Supersaturation over time at different values of $\lambda$ and static pressure of $p_{\scriptscriptstyle m}=1~\mathrm{MPa}$.}
\label{fig:Sovertime_alllamp1}
\end{figure}

\begin{figure}[ht!]
	\centering
	\includegraphics{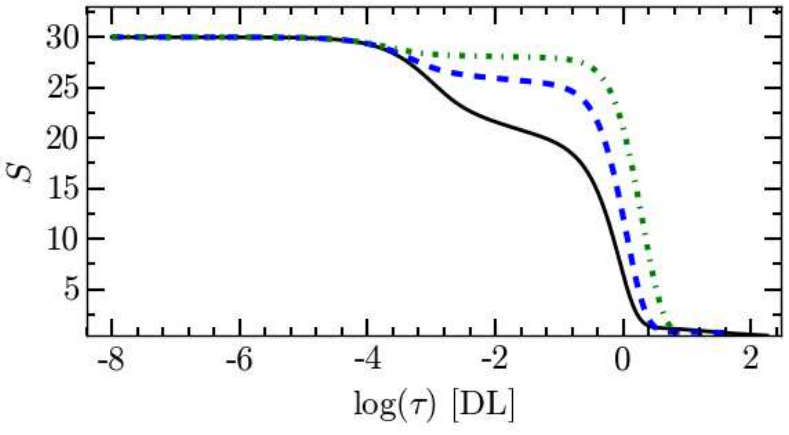}
	\caption{ Supersaturation over time at different values of $\lambda$ and static pressure of $p_{\scriptscriptstyle m}=100~\mathrm{MPa}$.The legend is the same as that of Figure \ref{fig:Sovertime_alllamp1}.}
	\label{fig:Sovertime_alllamp100}
\end{figure}

\begin{figure}[ht!]
	\centering
	\includegraphics{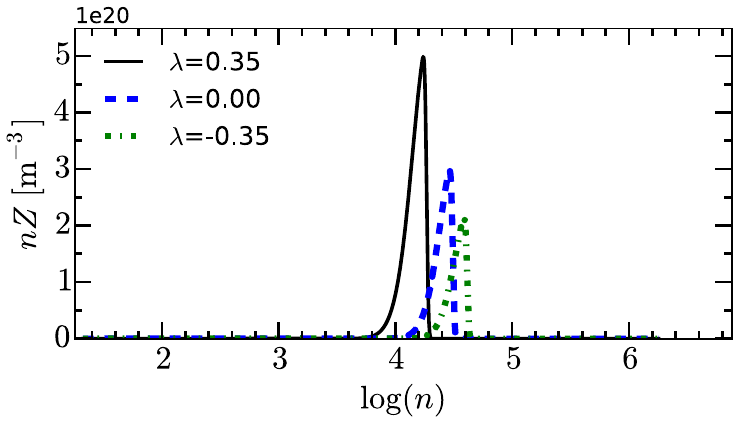}
	\caption{  The CSD at the end of nucleation stage ($\mathrm{log(}\tau\mathrm{)}=2$) at a static pressure of $50~\mathrm{MPa}$ and at different $\lambda$ values. }
	\label{fig:Z_n_Pac50MPa_Alllam}
\end{figure}

\begin{figure}[ht!]
	\centering
	\includegraphics{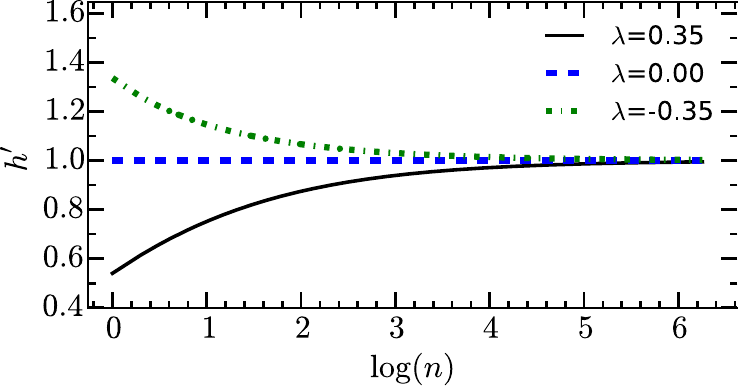}
	\caption{ $h^\prime$ at different $\lambda$ values for a range of cluster sizes. }
	\label{fig:hprime_n}
\end{figure}

\begin{figure}[ht!]
	\centering
	\subfigure[]{
		\includegraphics{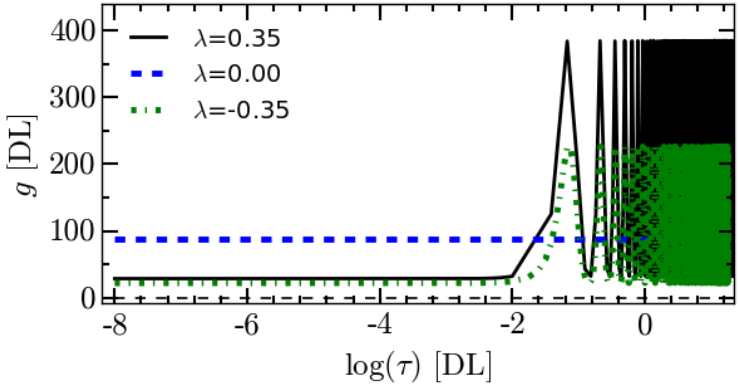}
	}
	\subfigure[]{
		\includegraphics{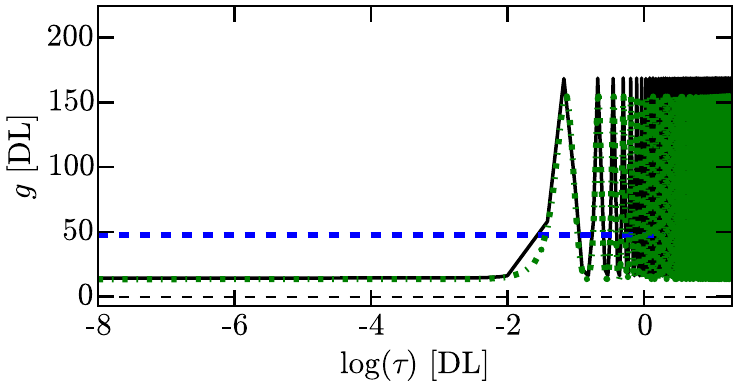}
	}
	\caption{ (a) Detachment frequency, Eq. \ref{eq:DetachFreq_Isotherm2}, for a supercritical cluster of size $n=70$ over time.  Parameters of acoustic wave are $f=2~\mathrm{MHz}$ and $p_{\scriptscriptstyle m}=50~\mathrm{MPa}$. (b) As in (a) but evaluated for a supercritical cluster of size $n=10^{4.5}$.}.
	\label{fig:govertime_AllLam}
\end{figure}

\begin{figure}[ht!]
	\centering
	\subfigure[]{
		\includegraphics{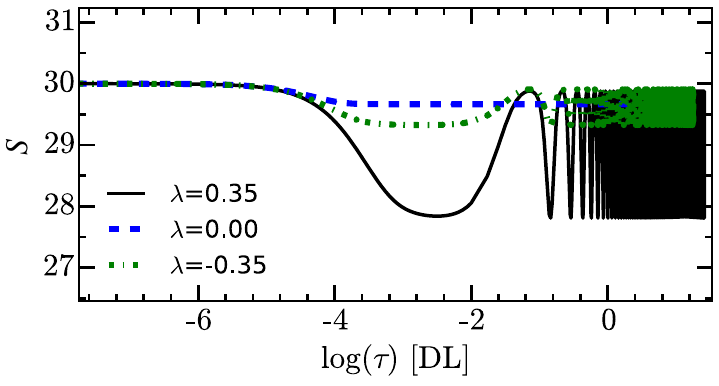}
	}
	\subfigure[]{
		\includegraphics{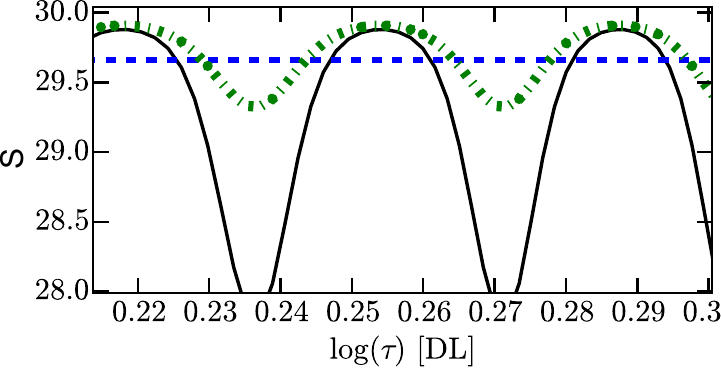}
	}
	\caption{ (a) Supersaturation over time where the old phase is exposed to an acoustic wave, the same acoustic wave parameters as in Figure \ref{fig:govertime_AllLam}, with different $\lambda$ values. (b) magnified about $\mathrm{log}(\tau)=0.25$. }
	\label{fig:Sovertime_AllLam}
\end{figure}

We demonstrated that by choosing a suitable $\lambda$ value, we could correctly predict the water droplet nucleation rate. To study the effect of an acoustic wave on crystal nucleation, we employed the size-independent $\lambda=0.35$. Here we perform a sensitivity analysis of the parameter $\lambda$ including the case of $\lambda=0$ representing the EDS cluster. Figures \ref{fig:Sovertime_alllamp1} and \ref{fig:Sovertime_alllamp100} show variation in supersaturation over time at two different pressure magnitudes and different $\lambda$ values. Given $k_\rho>0$, a negative $\lambda$ basically implies that the dividing surface is placed beyond the EDS.

The variation of supersaturation over time at different $\lambda$ and pressure magnitudes is depicted in Figures \ref{fig:Sovertime_alllamp1} and \ref{fig:Sovertime_alllamp100}. At a small static pressure magnitude, the effect of pressure on the thermodynamics and kinetics of nucleation is negligible and $\lambda$ influences the kinetics through $h$ and $h^\prime$ in the first two terms of Eq. \ref{eq:DetachFreq_Isotherm2}. We observed at low pressure magnitudes, a nucleation rate increase as $\lambda$ drops whereas at high magnitude static pressure, due to the role of the last term in Eq. \ref{eq:DetachFreq_Isotherm2}, the inverse trend was identified. This change in the nucleation rate influences the CSD at different $\lambda$ values. For instance, at $p_{\scriptscriptstyle m}=50~\mathrm{MPa}$ and at the end of nucleation stage, we see that the mean of the CSD is shifted towards a smaller $n$ (Figure \ref{fig:Z_n_Pac50MPa_Alllam}). This difference becomes more noticeable at higher pressure magnitudes.

In the case of pressure fluctuation with non-zero frequency, the effect of the location of dividing surface on nucleation is more clear.  Inspecting Eq. \ref{eq:DetachFreq_Isotherm2} shows that a non-EDS cluster can affect the kinetics of nucleation through the values of $n_n,~h$ and $h^\prime$. For the EDS cluster, $h$ and $h^\prime$ are constant and size-independent (equal to $0$ and $1$, respectively). However, for a non-EDS cluster these quantities are variable and size-dependent, see Figure \ref{fig:hprime_n}. This influences the pressure effect on the depletion rate and nucleation rate consequently, as shown in Figure \ref{fig:govertime_AllLam} for two different supercritical clusters.

The simulation results at  $p_{\scriptscriptstyle m}=50~\mathrm{MPa}$ and $f=2~\mathrm{MHz}$, shown in Figure \ref{fig:Sovertime_AllLam}, displays a variable supersaturation over time for both non-EDS cases whereas it is roughly non-oscillatory for the EDS cluster. This is particular to this combination of supersaturation and pressure magnitude as we observed a fluctuating supersaturation for the case of EDS clusters either at lower initial supersaturation or higher pressure magnitude. This is explained by the inverse relationship between pressure and supersaturation such that the pressure effect becomes more significant at lower supersaturations \cite{ford1991,kashchiev1995} and therefore imposes variation in detachment frequency and supersaturation eventually.

\section{Conclusion}\label{sec:Conc}

The Gibbs formalism is often employed to determine the thermodynamics of a phase transformation. This model assumes that the new phase forms a cluster of molecules separated from the old phase by a sharp, i.e. zero volume, interface phase. For large clusters, the deviations between the core new phase modelled with continuum properties and the real structure of the new phase are physically associated with the interface phase (and its excess free energy). However, this does not hold so readily for small clusters of the size of few molecules; for instance the density of the core new phase deviates from the bulk condensed phase density. \cite{ford1996} Nevertheless, we have shown that the Gibbs model can overcome some of these difficulties associated with the thermodynamics of small clusters if a non-EDS is utilised to define a cluster. For a given cluster size, moving a dividing surface essentially modifies the size of the core new phase and its thermodynamics. Furthermore, the specification of the dividing surface influences the excess Helmholtz free energy of the interface phase, given by $F_{s,1} ~=~ n_\sigma \Delta \mu+\Omega_\sigma$, and consequently the effective surface tension: see Eq. \ref{eq:GammaEff}. The dividing surface is the unphysical element of the model and its corresponding surface tension is defined to make the free energy of the interface phase independent of the location of the dividing surface.\cite{ford1996}  Derivations in the paper are valid for any dividing surface, including the EDS and surface of tension, and their associated size-dependent surface tension $\gamma(n)$. Computation of the excess free energy of the surface $F_{s,1} $ requires the knowledge about the size of interface phase $n_\sigma$ and the surface tension. Equation \ref{eq:Nsigma} is developed to calculate the size of the core new phase and the interface phase for any location of a dividing surface relative to the conventional EDS. Selecting $\gamma(n)$ requires a suitable model of the size-dependent surface tension but many of those available models often break down in the limit of small clusters. This issue becomes more significant in the case of sonocrystallisation process: the critical cluster size (for a condensed new phase) decreases as the pressure magnitude increases.

Therefore, we defined \emph{the new surface} which is identified as follows: i) this surface is characterised by the size-independent surface tension $\gamma_\infty$, and ii) the surface is positioned such that we obtain a reference excess free surface energy for the clusters. This was achieved by equating $\gamma_{\mathrm{\scriptscriptstyle eff}}$ (obtained from Eq. \ref{eq:GammaEff} when setting $\gamma(n)=\gamma_{\mathrm{\scriptscriptstyle \infty}}$) to the effective surface tension obtained from statistical mechanical simulations and solving for the parameter $\lambda$.  We showed that even a size-independent $\lambda$ and associated non-EDS clusters can reasonably well reproduce the excess free energy of different cluster sizes obtained from statistical mechanical simulations and successfully predict the kinetics of water droplet formation.

In addition, the effect of pressure variation on the cluster formation kinetics was studied. We demonstrated that this effect is cluster size-dependent. This is introduced by the term $\Delta n^\prime_{exc}$ in Eq. \ref{eq:DetachFreq_Cont2} which is illustrated in Figure \ref{fig:hprime_n} as well. In contrast to EDS clusters used in the CNT, the effect of pressure on the work of cluster formation and consequently the detachment rate varies with the size of non-EDS clusters (the work and detachment rate can be decreased or increased depending on the sign of $\lambda$) and tends towards the predictions of the CNT in the limit of large clusters. For an EDS cluster we have $h^\prime=1$ and therefore $\Delta n^\prime_{exc}=k_{\rho}$ which becomes negligible if the difference in molecular density of the old and new phases is small. This impairs the effect of pressure on the thermodynamics of phase transformation. In contrast, a non-EDS cluster gives a non-unity $h^\prime$, especially for a small cluster size, see Figure \ref{fig:hprime_n}, and the thermodynamic effect of a pressure variation can be more substantial. Additionally, the size of a condensed critical cluster inversely correlates to the pressure magnitude, see dashed black curves in Figures \ref{fig:Zsn_tau_n_Pac1MPa}-\ref{fig:Zsn_tau_n_Pac100MPa}. This together with the size-dependence of $h^\prime$ may explain some sonocrystallisation experimental observations revealing the improvement in a nucleation rate for a scenario with $\rho \approx \rho_n$ \cite{harzali2011} while the the conventional form of the CNT which uses EDS clusters is incapable of doing so.

With regard to the effect of pressure oscillation on a phase transformation, especially when the new phase is condensed and incompressible, pressure fluctuation can in general enhance or diminish the nucleation rate of the new phase by changing the nucleation barrier. Equations \ref{eq:WorkdiffCondFinal2} and \ref{eq:NucWorkDiff3} demonstrate the effect of pressure on the work of non-critical and critical cluster formation, respectively. We can see that for a denser new phase, pressure elevation reduces the nucleation barrier and consequently favours nucleation kinetics whereas pressure reduction increases the nucleation barrier and consequently lessens the probability of nucleation. The inverse trend happens for a new phase with $\Delta n_{exc}<0$. As a result, the nucleation rate goes up in a half cycle of acoustic waves but diminishes in the other half cycle. This is due to variation in detachment frequency with pressure oscillation as shown in Figure \ref{fig:govertime_AllLam} for a crystallisation process. In a half cycle, the detachment frequency is lower than the attachment frequency which leads to cluster growth whereas in the other half cycle detachment frequency becomes larger than attachment frequency which promotes the decay of the cluster. This leads to a time variable nucleation rate which alters the size distribution of supercritical clusters. The effect of variation in the static pressure on nucleation kinetics is binary, either enhancement or attenuation, however the acoustic wave produces both effects over a cycle. A precise experiment on the nucleation of solid helium from liquid helium conducted by Chavanne et. al. \cite{chavanne2001} illustrated a similar observation over a cycle of acoustic irradiation by a hemi-spherical focused ultrasound transducer.

The thermodynamics and kinetics of a phase transformation in a closed system exposed to an acoustic field and governed only by the aggregative mechanism has been investigated. Adding the non-aggregative effect of an acoustic wave into the developments made in this work and solving the coupled problem will be the subject of a forthcoming paper.

\begin{acknowledgments}
This work was supported by the EPSRC [grant number EP/I031480/1].
\end{acknowledgments}

\appendix

\section{Number of excess molecules } \label{sec:AppxNsigma}

The number of molecules in the interface phase of a cluster with an arbitrary shape can be determined by Eq. \ref{eq:InterfaceNoFinal}. For a spherical cluster this equation reads

\begin{eqnarray}
\label{eq:InterfaceNoSpher}
& n_\sigma = (\rho_n  - \rho) (V_n^{E} - V_n) = \dfrac{4 \pi}{3}(\rho_n  - \rho) ({R_\sigma^{E}}^3 - R_\sigma^3) \nonumber \\
& = \dfrac{4 \pi}{3}(\rho_n  - \rho) (R_\sigma^{E} - R_\sigma)({R_\sigma^{E}}^2 + R_\sigma R_\sigma^{E} + R_\sigma^2),
\end{eqnarray}

\noindent where $R_\sigma^E $ is the radius of a cluster defined by the equimolar surface. Considering $\delta = R_\sigma^E - R_\sigma$, it follows

\begin{eqnarray}
\label{eq:InterfaceNoSpher2}
 & n_\sigma =  \dfrac{4 \pi}{3}(\rho_n  - \rho) \delta ({R_\sigma^{E}}^2 +  {R_\sigma^{E}}^2 - R_\sigma^E \delta + {R_\sigma^{E}}^2 + \delta^2 - 2R_\sigma^E \delta)  \nonumber \\
& = \dfrac{4 \pi}{3} \delta (\rho_n  - \rho) (3 {R_\sigma^{E}}^2 - 3  R_\sigma^E \delta + \delta^2).
\end{eqnarray}

\noindent  Substituting $R_\sigma^E = \left( \tfrac{3\nu_n}{4 \pi} \right)^{\textsc{\tiny{1/3}}} n_e^{\textsc{\tiny{1/3}}} = R_0 n_e^{\textsc{\tiny{1/3}}}$ where $R_0=\left( \tfrac{3\nu_n}{4 \pi} \right)^{\textsc{\tiny{1/3}}}$ is the average intermolecular distance in the bulk of new phase, in the above equation yields

\begin{eqnarray}
\label{eq:InterfaceNoSpher3}
& n_\sigma =  4 \pi  R_0^2 \delta (\rho_n  - \rho) \left(   n_e^{\tfrac{2}{3}}  -  \dfrac{\delta}{R_0}  n_e^{\tfrac{1}{3}} + \dfrac{\delta^2}{3R_0^2}  \right)  \nonumber \\
& = k_\sigma \left(   n_e^\beta  -  \dfrac{\delta}{R_0}  n_e^{1-\beta} + \dfrac{\delta^2}{3R_0^2}  \right),
\end{eqnarray}

\noindent where  $k_\sigma =  4 \pi  R_0^2 \delta (\rho_n  - \rho) $ and $\beta=\tfrac{2}{3}$. We can write $k_\sigma $ as follows $k_\sigma =  k_\rho 3 \lambda$ where $k_\rho = 1-\tfrac{\rho}{\rho_n}$ and $\lambda = \tfrac{\delta}{R_0}$ are dimensionless quantities.

For a cubic cluster with the length of $l_E$, we have $V_n^E=l_E^3=n_e \nu_n$. Plugging this in Eq. \ref{eq:InterfaceNoFinal} and after some algebra we arrive in

\begin{equation}
\label{eq:InterfaceNoCube}
n_\sigma =  k_\sigma \left(   n_e^\beta  -  \dfrac{\lambda}{S_f}  n_e^{1-\beta} + \dfrac{\lambda^2}{3 S_f^2}  \right),
\end{equation}

\noindent where $k_\sigma =  \tfrac{1}{S_f} k_\rho 3 \lambda $  and $S_f = \left( \tfrac{4 \pi}{3} \right) ^{\textsc{\tiny{1/3}}}$ is the shape factor and $\beta=\tfrac{2}{3}$. Therefore we can use the formula in the form of Eq. \ref{eq:InterfaceNoCube} for both spherical and cubic clusters with shape factors of $S_f=1$ and $S_f = \left( \tfrac{4 \pi}{3} \right) ^{\textsc{\tiny{1/3}}}$ respectively.

Having determined the number of excess molecules and utilising Eq. \ref{eq:nExcess}, the number of molecules in the new phase (core) is obtained by

\begin{equation}
\label{eq:CoreNo}
n_n = n_e - \dfrac{n_\sigma}{k_\rho} = n_e - \dfrac{3 \lambda}{S_f} \left(   n_e^\beta  -  \dfrac{\lambda}{S_f}  n_e^{1-\beta} + \dfrac{\lambda^2}{3 S_f^2}   \right).
\end{equation}

\noindent Therefore, in general we can write

\begin{eqnarray}
\label{eq:InterfaceNoGen}
&& n_\sigma(n_e) = k_\rho \mathcal{F}(n_e), \\
&& n_n(n_e) = n_e-\mathcal{F}(n_e),
\end{eqnarray}

\noindent where $\mathcal{F}(n_e) =  \tfrac{3 \lambda}{S_f} \left(   n_e^\beta  -  \tfrac{\lambda}{S_f} n_e^{1-\beta} + \tfrac{\lambda^2}{3 S_f^2}  \right) $.

If $\lambda \ll 1$, i.e. $\delta \ll R_0$, the number of molecules in the interface and new phase can be approximated with second order error ($\mathcal{O}(\lambda^2)$) as follows

\begin{eqnarray}
\label{eq:InterfaceNoAppr}
 && n_\sigma (n_e) =  k_\rho \tfrac{3 \lambda}{S_f}  n_e^\beta + \mathcal{O}(\lambda^2), \\
 && n_n(n_e) = n_e - \tfrac{3 \lambda}{S_f} n_e^\beta + \mathcal{O}(\lambda^2).
\end{eqnarray}

These equations give $n_n$ and $n_\sigma$ as a function of $n_e$ which is the size of the EDS-defined cluster. We are, however, interested in determining these quantities and the size of cluster as the function of either the core size or the number of molecules in the interface. In this regard, we start with Eq. \ref{eq:CoreNo} and solve it for $n_e$ while employing $\beta=\tfrac{2}{3}$ as follows

\begin{equation}
\label{eq:NeNo}
n_n = n_e - \dfrac{3 \lambda}{S_f} \left(   n_e^{\tfrac{2}{3}}  -  \dfrac{\lambda}{S_f}  n_e^{\tfrac{1}{3}} + \dfrac{\lambda^2}{3 S_f^2}   \right) = \left( n_e^{\tfrac{1}{3}} - \dfrac{\lambda}{S_f}  \right)^3,
\end{equation}

\noindent which gives

\begin{equation}
\label{eq:NeNo2}
n_e = \left( n_n^{\tfrac{1}{3}} + \dfrac{\lambda}{S_f}  \right)^3,
\end{equation}

\noindent by substituting this relationship in Eq. \ref{eq:InterfaceNoCube}, we obtain $n_\sigma$ and the cluster size in the following format

\begin{eqnarray}
\label{eq:NCluster}
&& n_\sigma(n_n) = k_\rho \mathcal{G}(n_n), \\
&& n(n_n) = n_n + n_\sigma(n_n) = n_n + k_\rho \mathcal{G}(n_n),
\end{eqnarray}

\noindent where $\mathcal{G}(n_n) =  \tfrac{3 \lambda}{S_f} \left(    n_n^{\tfrac{2}{3}}  +  \tfrac{\lambda}{S_f}  n_n^{\tfrac{1}{3}} + \tfrac{\lambda^2}{3 S_f^2}  \right) $.

For the case of a condensed new phase, if the dividing surface is placed beyond the EDS; this gives $\lambda<0$ and subsequently $\mathcal{G}(n_n)<0$. On the other hand, if the surface is enclosed in the EDS, we have $\lambda>0, \mathcal{G}(n_n)>0$. Finally, we need to determine the derivatives of $n_n$ and $n_\sigma$ with respect to $n$ as they are required in Eq. \ref{eq:Work9}. Using the last two equations, we have

\begin{eqnarray}
\label{eq:DiffNsigma}
&& h(n_n)~=~\dfrac{d n_\sigma}{d n} ~=~ \dfrac{d n_\sigma}{d n_n} \dfrac{d n_n}{d n} =\nonumber \\
&&  \left[ k_\rho \dfrac{3 \lambda}{S_f} \left(  \dfrac{2}{3}  n_n^{- \tfrac{1}{3}}  +  \dfrac{1}{3} \dfrac{\lambda}{S_f}  n_n^{- \tfrac{2}{3}}  \right) \right]  \dfrac{d n_n}{d n},
\end{eqnarray}

\noindent furthermore

\begin{equation}
\label{eq:DiffNn}
\dfrac{d n}{d n_n} ~=~ 1 +   k_\rho \dfrac{3 \lambda}{S_f} \left(  \dfrac{2}{3}  n_n^{- \tfrac{1}{3}}  +  \dfrac{1}{3} \dfrac{\lambda}{S_f}  n_n^{- \tfrac{2}{3}}  \right),
\end{equation}

\noindent which gives

\begin{equation}
\label{eq:DiffNn1}
h^{'}(n_n) = \dfrac{d n_n}{d n} ~=~ \frac{1} {1 +   k_\rho \dfrac{3 \lambda}{S_f} \left(  \dfrac{2}{3}  n_n^{- \tfrac{1}{3}}  +  \dfrac{1}{3} \dfrac{\lambda}{S_f}  n_n^{- \tfrac{2}{3}}  \right) },
\end{equation}

\noindent eventually plugging Eq. \ref{eq:DiffNn1} into Eq. \ref{eq:DiffNsigma} gives

\begin{equation}
\label{eq:DiffNsigma1}
h(n_n) = \dfrac{d n_\sigma}{d n} ~=~  \frac{k_\rho \dfrac{3 \lambda}{S_f} \left(  \dfrac{2}{3}  n_n^{- \tfrac{1}{3}}  +  \dfrac{1}{3} \dfrac{\lambda}{S_f}  n_n^{- \tfrac{2}{3}}  \right) } {1 +   k_\rho \dfrac{3 \lambda}{S_f} \left(  \dfrac{2}{3}  n_n^{- \tfrac{1}{3}}  +  \dfrac{1}{3} \dfrac{\lambda}{S_f}  n_n^{- \tfrac{2}{3}}  \right) } .
\end{equation}

\section{Chemical potential of new phase } \label{sec:AppxChemPot}
The Gibbs-Duhem equation when temperature and composition for the new phase are kept constant is as follows: \cite{guggenheim1985} $d\mu_n(p) = \nu_n dp$. Integrating both sides given that the new phase is condensed, we arrive at

\begin{eqnarray}
&& \int\limits_{p}^{p_n} d\mu_n(p) =  \int\limits_{p}^{p_n} \nu_n dp \nonumber \\
&& \mu_n(p_n) - \mu_n(p) = \nu_n \left( p_n - p \right).
\label{eq:ChemPotDiff}
\end{eqnarray}

\noindent This equation can be rearranged as $\mu_n(p) = \mu_n(p_n)$ $ - \nu_n \left( p_n - p \right)$. Differentiating both sides of this equation while the partial molar volume is kept constant yields: $d \mu_n(p) = d\mu_n(p_n) - \nu_n \left(d p_n - dp \right)$. The Gibbs-Duhem relationship for the new phase also reads: $d\mu_n(p_n) = -s_ndT + \nu_n dp_n$. Combining the last two equations gives the differential form of the chemical potential of the new phase at pressure $p$ of old phase which was used in the text.

In addition, evaluating Eq. \ref{eq:ChemPotDiff} at the pressure of a critical cluster $p^*_n$ gives:

\begin{eqnarray}
&& \mu_n(p_n) - \mu_n(p^*_n) = \nu_n \left( p_n - p^*_n \right).
\label{eq:ChemPotDiffCritical}
\end{eqnarray}

The equilibrium condition for a critical cluster yields $\mu_n(p^*_n)=\mu(p)$. Substituting this in Eq. \ref{eq:ChemPotDiffCritical} and using Eq. \ref{eq:ChemPotDiff} gives

\begin{equation}
p_n-p=\rho_n\Delta \mu + (p_n-p_n^*).
\end{equation}

\section{Discrete form of monomer detachment frequency } \label{sec:AppxDisctForm}

In the discrete representation of the cluster formation work, Eq. \ref{eq:DetachFreq_Disc} is used to determine detachment frequency. So, $\Delta \Omega_n(~)-\Delta \Omega_{n-1}(~)=\int \left( d\Delta \Omega_n(~)~-~d\Delta \Omega_{n-1}(~)\right)$ should be determined where $(~)$ denotes the dependency of work on all other parameters, e.g. pressure, temperature and composition, which is omitted here to avoid long relations. The two terms of the integrand are obtained with the aid of Eq.  \ref{eq:WorkdiffCondFinal}. The integrand then becomes

\begin{eqnarray}
\label{eq:Work12_Disc}
&& d\Delta \Omega_n(~)-d\Delta \Omega_{n-1}(~)=-[ -s + s_n(n_{n,n} - n_{n,n-1}) + \nonumber \\
&& s_\sigma(n_{\sigma,n} - n_{\sigma,n-1})]dT - \left[ \nu - \nu_n(n_{n,n}-n_{n,n-1})\right]dp.  \nonumber \\
\end{eqnarray}

\noindent In order to simplify this equation, we need to determine $n_{n,n} - n_{n,n-1}$ and $n_{\sigma,n} - n_{\sigma,n-1}$. Employing Eq. \ref{eq:NCluster} gives

\begin{eqnarray}
\label{eq:NClusterDisc1}
&& n = n_{n,n} + n_{\sigma,n} = n_{n,n} + k_\rho \mathcal{G}(n_{n,n})   \nonumber \\
&& n-1 = n_{n,n-1} + n_{\sigma,n-1} = n_{n,n-1} + k_\rho \mathcal{G}(n_{n,n-1}), \nonumber \\
\end{eqnarray}

\noindent which follows

\begin{eqnarray}
\label{eq:NClusterDisc2}
&& 1 = n_{n,n} - n_{n,n-1} + n_{\sigma,n} - n_{\sigma,n-1}  = \nonumber \\
&& n_{n,n}- n_{n,n-1}  + k_\rho 3 \lambda \left( n_{n,n}^{\tfrac{2}{3}} - n_{n,n-1}^{\tfrac{2}{3}}  + \lambda (n_{n,n}^{\tfrac{1}{3}} - n_{n,n-1}^{\tfrac{1}{3}} )  \right). \nonumber \\
\end{eqnarray}

\noindent Defining the new variable $X=n_{n,n}- n_{n,n-1}$, the above equation reads

\begin{eqnarray}
\label{eq:NClusterDisc3}
&& 1 = X + n_{\sigma,n} - n_{\sigma,n-1}  = \nonumber \\
&& X  + k_\rho 3 \lambda \left( n_{n,n}^{\tfrac{2}{3}} - (n_{n,n}-X)^{\tfrac{2}{3}}  + \lambda (n_{n,n}^{\tfrac{1}{3}} - (n_{n,n}-X)^{\tfrac{1}{3}} )  \right). \nonumber \\
\end{eqnarray}

\noindent Approximating the term $(n_{n,n}-X)^\beta$ by its second order truncated binomial expansion gives

\begin{eqnarray}
\label{eq:BinomTrunc}
&& (n_{n,n}-X)^\beta \approx n_{n,n}^\beta-\beta n_{n,n}^{\beta-1} X.
\end{eqnarray}

\noindent Given $\beta<1$ this approximation introduces negligible error. Substituting this approximation in Eqs. \ref{eq:NClusterDisc3} yields

\begin{equation}
\label{eq:NClusterDisc4}
1 = X  + k_\rho 3 \lambda X \left(  \frac{2}{3}n_{n,n}^{-\tfrac{1}{3}}  + \frac{\lambda}{3}n_{n,n}^{-\tfrac{2}{3}}   \right),
\end{equation}

\noindent from which follows

\begin{equation}
\label{eq:NClusterDisc5}
X = \frac{1}{1  + k_\rho 3 \lambda \left(  \frac{2}{3}n_{n,n}^{-\tfrac{1}{3}}  + \frac{\lambda}{3}n_{n,n}^{-\tfrac{2}{3}}   \right) }.
\end{equation}

\noindent This is the same as $h^{'}(n_n)$ in Eq. \ref{eq:DiffNn1} and \ref{eq:Work9}. Plugging $X$ into Eq. \ref{eq:NClusterDisc3} we calculate $n_{\sigma,n} - n_{\sigma,n-1} $ which reads the same as $h(n_n)$ in Eq. \ref{eq:DiffNsigma1} and \ref{eq:Work9}. Performing the integration gives exactly the same results already achieved for the case of variation in work as a function of continuous $n$ shown in Eq. \ref{eq:Work9}. Considering the second order binomial truncation is applied, then all the equations derived previously to calculate detachment frequency are valid and can be used for the discrete representation of $n$, too.


\begin{thebibliography}{56}%
\makeatletter
\providecommand \@ifxundefined [1]{%
 \@ifx{#1\undefined}
}%
\providecommand \@ifnum [1]{%
 \ifnum #1\expandafter \@firstoftwo
 \else \expandafter \@secondoftwo
 \fi
}%
\providecommand \@ifx [1]{%
 \ifx #1\expandafter \@firstoftwo
 \else \expandafter \@secondoftwo
 \fi
}%
\providecommand \natexlab [1]{#1}%
\providecommand \enquote  [1]{``#1''}%
\providecommand \bibnamefont  [1]{#1}%
\providecommand \bibfnamefont [1]{#1}%
\providecommand \citenamefont [1]{#1}%
\providecommand \href@noop [0]{\@secondoftwo}%
\providecommand \href [0]{\begingroup \@sanitize@url \@href}%
\providecommand \@href[1]{\@@startlink{#1}\@@href}%
\providecommand \@@href[1]{\endgroup#1\@@endlink}%
\providecommand \@sanitize@url [0]{\catcode `\\12\catcode `\$12\catcode
  `\&12\catcode `\#12\catcode `\^12\catcode `\_12\catcode `\%12\relax}%
\providecommand \@@startlink[1]{}%
\providecommand \@@endlink[0]{}%
\providecommand \url  [0]{\begingroup\@sanitize@url \@url }%
\providecommand \@url [1]{\endgroup\@href {#1}{\urlprefix }}%
\providecommand \urlprefix  [0]{URL }%
\providecommand \Eprint [0]{\href }%
\providecommand \doibase [0]{http://dx.doi.org/}%
\providecommand \selectlanguage [0]{\@gobble}%
\providecommand \bibinfo  [0]{\@secondoftwo}%
\providecommand \bibfield  [0]{\@secondoftwo}%
\providecommand \translation [1]{[#1]}%
\providecommand \BibitemOpen [0]{}%
\providecommand \bibitemStop [0]{}%
\providecommand \bibitemNoStop [0]{.\EOS\space}%
\providecommand \EOS [0]{\spacefactor3000\relax}%
\providecommand \BibitemShut  [1]{\csname bibitem#1\endcsname}%
\let\auto@bib@innerbib\@empty
\bibitem [{\citenamefont {Neppiras}\ and\ \citenamefont
  {Noltingk}(1951)}]{neppiras1951}%
  \BibitemOpen
  \bibfield  {author} {\bibinfo {author} {\bibfnamefont {E.}~\bibnamefont
  {Neppiras}}\ and\ \bibinfo {author} {\bibfnamefont {B.}~\bibnamefont
  {Noltingk}},\ }\bibfield  {title} {\enquote {\bibinfo {title} {Cavitation
  produced by ultrasonics: theoretical conditions for the onset of
  cavitation},}\ }\href@noop {} {\bibfield  {journal} {\bibinfo  {journal}
  {Proceedings of the Physical Society. Section B}\ }\textbf {\bibinfo {volume}
  {64}},\ \bibinfo {pages} {1032} (\bibinfo {year} {1951})}\BibitemShut
  {NoStop}%
\bibitem [{\citenamefont {Blander}\ and\ \citenamefont
  {Katz}(1975)}]{blander1975}%
  \BibitemOpen
  \bibfield  {author} {\bibinfo {author} {\bibfnamefont {M.}~\bibnamefont
  {Blander}}\ and\ \bibinfo {author} {\bibfnamefont {J.~L.}\ \bibnamefont
  {Katz}},\ }\bibfield  {title} {\enquote {\bibinfo {title} {Bubble nucleation
  in liquids},}\ }\href@noop {} {\bibfield  {journal} {\bibinfo  {journal}
  {AIChE Journal}\ }\textbf {\bibinfo {volume} {21}},\ \bibinfo {pages}
  {833--848} (\bibinfo {year} {1975})}\BibitemShut {NoStop}%
\bibitem [{\citenamefont {Akulichev}(1982)}]{akulichev1982}%
  \BibitemOpen
  \bibfield  {author} {\bibinfo {author} {\bibfnamefont {V.~A.}\ \bibnamefont
  {Akulichev}},\ }\bibfield  {title} {\enquote {\bibinfo {title} {Acoustic
  cavitation in cryogenic and boiling liquids},}\ }in\ \href@noop {} {\emph
  {\bibinfo {booktitle} {Mechanics and Physics of Bubbles in Liquids}}}\
  (\bibinfo  {publisher} {Springer},\ \bibinfo {year} {1982})\ pp.\ \bibinfo
  {pages} {55--67}\BibitemShut {NoStop}%
\bibitem [{\citenamefont {Baidakov}, \citenamefont {Kaverin},\ and\
  \citenamefont {Skripov}(1981)}]{baidakov1981}%
  \BibitemOpen
  \bibfield  {author} {\bibinfo {author} {\bibfnamefont {V.}~\bibnamefont
  {Baidakov}}, \bibinfo {author} {\bibfnamefont {A.}~\bibnamefont {Kaverin}}, \
  and\ \bibinfo {author} {\bibfnamefont {V.}~\bibnamefont {Skripov}},\
  }\bibfield  {title} {\enquote {\bibinfo {title} {Acoustic cavitation in
  superheated liquids},}\ }\href@noop {} {\bibfield  {journal} {\bibinfo
  {journal} {Akusticheskii Zhurnal}\ }\textbf {\bibinfo {volume} {27}},\
  \bibinfo {pages} {697--703} (\bibinfo {year} {1981})}\BibitemShut {NoStop}%
\bibitem [{\citenamefont {Kapustin}(1963)}]{kapustin1963}%
  \BibitemOpen
  \bibfield  {author} {\bibinfo {author} {\bibfnamefont {A.}~\bibnamefont
  {Kapustin}},\ }\href@noop {} {\emph {\bibinfo {title} {The effects of
  ultrasound on the kinetics of crystallization}}}\ (\bibinfo  {publisher}
  {Consultants Bureau, New York},\ \bibinfo {year} {1963})\BibitemShut
  {NoStop}%
\bibitem [{\citenamefont {Akulichev}\ and\ \citenamefont
  {Bulanov}(1983)}]{akulichev1983}%
  \BibitemOpen
  \bibfield  {author} {\bibinfo {author} {\bibfnamefont {V.}~\bibnamefont
  {Akulichev}}\ and\ \bibinfo {author} {\bibfnamefont {V.}~\bibnamefont
  {Bulanov}},\ }\bibfield  {title} {\enquote {\bibinfo {title} {Crystallization
  nuclei in liquid in a sound field},}\ }\href@noop {} {\bibfield  {journal}
  {\bibinfo  {journal} {International Journal of Heat and Mass Transfer}\
  }\textbf {\bibinfo {volume} {26}},\ \bibinfo {pages} {289 -- 300} (\bibinfo
  {year} {1983})}\BibitemShut {NoStop}%
\bibitem [{\citenamefont {Arakelyan}(1987)}]{arakelyan1987}%
  \BibitemOpen
  \bibfield  {author} {\bibinfo {author} {\bibfnamefont {V.}~\bibnamefont
  {Arakelyan}},\ }\bibfield  {title} {\enquote {\bibinfo {title} {Effect of
  ultrasound on crystal growth from melt and solution},}\ }\href@noop {}
  {\bibfield  {journal} {\bibinfo  {journal} {Acta Physica Hungarica}\ }\textbf
  {\bibinfo {volume} {61}},\ \bibinfo {pages} {185--187} (\bibinfo {year}
  {1987})}\BibitemShut {NoStop}%
\bibitem [{\citenamefont {Chavanne}, \citenamefont {Balibar},\ and\
  \citenamefont {Caupin}(2001)}]{chavanne2001}%
  \BibitemOpen
  \bibfield  {author} {\bibinfo {author} {\bibfnamefont {X.}~\bibnamefont
  {Chavanne}}, \bibinfo {author} {\bibfnamefont {S.}~\bibnamefont {Balibar}}, \
  and\ \bibinfo {author} {\bibfnamefont {F.}~\bibnamefont {Caupin}},\
  }\bibfield  {title} {\enquote {\bibinfo {title} {Acoustic crystallization and
  heterogeneous nucleation},}\ }\href@noop {} {\bibfield  {journal} {\bibinfo
  {journal} {Physical Review Letters}\ }\textbf {\bibinfo {volume} {86}},\
  \bibinfo {pages} {5506--5509} (\bibinfo {year} {2001})}\BibitemShut {NoStop}%
\bibitem [{\citenamefont {Hem}(1967)}]{hem1967}%
  \BibitemOpen
  \bibfield  {author} {\bibinfo {author} {\bibfnamefont {S.~L.}\ \bibnamefont
  {Hem}},\ }\bibfield  {title} {\enquote {\bibinfo {title} {The effect of
  ultrasonic vibrations on crystallization processes},}\ }\href@noop {}
  {\bibfield  {journal} {\bibinfo  {journal} {Ultrasonics}\ }\textbf {\bibinfo
  {volume} {5}},\ \bibinfo {pages} {202--207} (\bibinfo {year}
  {1967})}\BibitemShut {NoStop}%
\bibitem [{\citenamefont {Ruecroft}\ \emph {et~al.}(2005)\citenamefont
  {Ruecroft}, \citenamefont {Hipkiss}, \citenamefont {Ly}, \citenamefont
  {Maxted},\ and\ \citenamefont {Cains}}]{ruecroft2005}%
  \BibitemOpen
  \bibfield  {author} {\bibinfo {author} {\bibfnamefont {G.}~\bibnamefont
  {Ruecroft}}, \bibinfo {author} {\bibfnamefont {D.}~\bibnamefont {Hipkiss}},
  \bibinfo {author} {\bibfnamefont {T.}~\bibnamefont {Ly}}, \bibinfo {author}
  {\bibfnamefont {N.}~\bibnamefont {Maxted}}, \ and\ \bibinfo {author}
  {\bibfnamefont {P.}~\bibnamefont {Cains}},\ }\bibfield  {title} {\enquote
  {\bibinfo {title} {Sonocrystallization: the use of ultrasound for improved
  industrial crystallization},}\ }\href@noop {} {\bibfield  {journal} {\bibinfo
   {journal} {Organic Process Research and Development}\ }\textbf {\bibinfo
  {volume} {9}},\ \bibinfo {pages} {923--932} (\bibinfo {year}
  {2005})}\BibitemShut {NoStop}%
\bibitem [{\citenamefont {Suslick}(1988)}]{suslick1988}%
  \BibitemOpen
  \bibfield  {author} {\bibinfo {author} {\bibfnamefont {K.~S.}\ \bibnamefont
  {Suslick}},\ }\href@noop {} {\emph {\bibinfo {title} {Ultrasound: its
  chemical, physical, and biological effects}}}\ (\bibinfo  {publisher} {VCH
  Publishers: New York},\ \bibinfo {year} {1988})\BibitemShut {NoStop}%
\bibitem [{\citenamefont {Mazhul}(1963)}]{mazhul1954}%
  \BibitemOpen
  \bibfield  {author} {\bibinfo {author} {\bibfnamefont {M.~M.}\ \bibnamefont
  {Mazhul}},\ }\bibfield  {title} {\enquote {\bibinfo {title} {Cavitation
  phenomena and the formation of crystal nuclei},}\ }\href@noop {} {\bibfield
  {journal} {\bibinfo  {journal} {PhD Dissertation}\ } (\bibinfo {year}
  {1963})}\BibitemShut {NoStop}%
\bibitem [{\citenamefont {Flint}\ and\ \citenamefont
  {Suslick}(1991)}]{flint1991}%
  \BibitemOpen
  \bibfield  {author} {\bibinfo {author} {\bibfnamefont {E.~B.}\ \bibnamefont
  {Flint}}\ and\ \bibinfo {author} {\bibfnamefont {K.~S.}\ \bibnamefont
  {Suslick}},\ }\bibfield  {title} {\enquote {\bibinfo {title} {The temperature
  of cavitation},}\ }\href@noop {} {\bibfield  {journal} {\bibinfo  {journal}
  {Science}\ }\textbf {\bibinfo {volume} {253}},\ \bibinfo {pages} {1397--1399}
  (\bibinfo {year} {1991})}\BibitemShut {NoStop}%
\bibitem [{\citenamefont {Suslick}\ and\ \citenamefont
  {Flannigan}(2008)}]{suslick2008}%
  \BibitemOpen
  \bibfield  {author} {\bibinfo {author} {\bibfnamefont {K.~S.}\ \bibnamefont
  {Suslick}}\ and\ \bibinfo {author} {\bibfnamefont {D.~J.}\ \bibnamefont
  {Flannigan}},\ }\bibfield  {title} {\enquote {\bibinfo {title} {Inside a
  collapsing bubble: sonoluminescence and the conditions during cavitation},}\
  }\href@noop {} {\bibfield  {journal} {\bibinfo  {journal} {Annual Review of
  Physical Chemistry}\ }\textbf {\bibinfo {volume} {59}},\ \bibinfo {pages}
  {659--683} (\bibinfo {year} {2008})}\BibitemShut {NoStop}%
\bibitem [{\citenamefont {Akhatov}\ \emph {et~al.}(2001)\citenamefont
  {Akhatov}, \citenamefont {Lindau}, \citenamefont {Topolnikov}, \citenamefont
  {Mettin}, \citenamefont {Vakhitova},\ and\ \citenamefont
  {Lauterborn}}]{akhatov2001}%
  \BibitemOpen
  \bibfield  {author} {\bibinfo {author} {\bibfnamefont {I.}~\bibnamefont
  {Akhatov}}, \bibinfo {author} {\bibfnamefont {O.}~\bibnamefont {Lindau}},
  \bibinfo {author} {\bibfnamefont {A.}~\bibnamefont {Topolnikov}}, \bibinfo
  {author} {\bibfnamefont {R.}~\bibnamefont {Mettin}}, \bibinfo {author}
  {\bibfnamefont {N.}~\bibnamefont {Vakhitova}}, \ and\ \bibinfo {author}
  {\bibfnamefont {W.}~\bibnamefont {Lauterborn}},\ }\bibfield  {title}
  {\enquote {\bibinfo {title} {Collapse and rebound of a laser-induced
  cavitation bubble},}\ }\href@noop {} {\bibfield  {journal} {\bibinfo
  {journal} {Physics of Fluids}\ }\textbf {\bibinfo {volume} {13}},\ \bibinfo
  {pages} {2805--2819} (\bibinfo {year} {2001})}\BibitemShut {NoStop}%
\bibitem [{\citenamefont {Ohl}\ \emph {et~al.}(1999)\citenamefont {Ohl},
  \citenamefont {Kurz}, \citenamefont {Geisler}, \citenamefont {Lindau},\ and\
  \citenamefont {Lauterborn}}]{ohl1999}%
  \BibitemOpen
  \bibfield  {author} {\bibinfo {author} {\bibfnamefont {C.~D.}\ \bibnamefont
  {Ohl}}, \bibinfo {author} {\bibfnamefont {T.}~\bibnamefont {Kurz}}, \bibinfo
  {author} {\bibfnamefont {R.}~\bibnamefont {Geisler}}, \bibinfo {author}
  {\bibfnamefont {O.}~\bibnamefont {Lindau}}, \ and\ \bibinfo {author}
  {\bibfnamefont {W.}~\bibnamefont {Lauterborn}},\ }\bibfield  {title}
  {\enquote {\bibinfo {title} {Bubble dynamics, shock waves and
  sonoluminescence},}\ }\href@noop {} {\bibfield  {journal} {\bibinfo
  {journal} {Philosophical Transactions of the Royal Society of London. Series
  A: Mathematical, Physical and Engineering Sciences}\ }\textbf {\bibinfo
  {volume} {357}},\ \bibinfo {pages} {269--294} (\bibinfo {year}
  {1999})}\BibitemShut {NoStop}%
\bibitem [{\citenamefont {Suslick}\ and\ \citenamefont
  {Price}(1999)}]{suslick1999}%
  \BibitemOpen
  \bibfield  {author} {\bibinfo {author} {\bibfnamefont {K.~S.}\ \bibnamefont
  {Suslick}}\ and\ \bibinfo {author} {\bibfnamefont {G.~J.}\ \bibnamefont
  {Price}},\ }\bibfield  {title} {\enquote {\bibinfo {title} {Application of
  ultrasound to materials chemistry},}\ }\href@noop {} {\bibfield  {journal}
  {\bibinfo  {journal} {Annual Review of Materials Science}\ }\textbf {\bibinfo
  {volume} {29}},\ \bibinfo {pages} {295--326} (\bibinfo {year}
  {1999})}\BibitemShut {NoStop}%
\bibitem [{\citenamefont {Ford}(1992)}]{ford1991}%
  \BibitemOpen
  \bibfield  {author} {\bibinfo {author} {\bibfnamefont {I.~J.}\ \bibnamefont
  {Ford}},\ }\bibfield  {title} {\enquote {\bibinfo {title} {Imperfect
  vapour-gas mixtures and homogeneous nucleation},}\ }\href@noop {} {\bibfield
  {journal} {\bibinfo  {journal} {Journal of Aerosol Science}\ }\textbf
  {\bibinfo {volume} {23}},\ \bibinfo {pages} {447 -- 455} (\bibinfo {year}
  {1992})}\BibitemShut {NoStop}%
\bibitem [{\citenamefont {Kashchiev}\ and\ \citenamefont
  {Van~Rosmalen}(1995)}]{kashchiev1995}%
  \BibitemOpen
  \bibfield  {author} {\bibinfo {author} {\bibfnamefont {D.}~\bibnamefont
  {Kashchiev}}\ and\ \bibinfo {author} {\bibfnamefont {G.~M.}\ \bibnamefont
  {Van~Rosmalen}},\ }\bibfield  {title} {\enquote {\bibinfo {title} {Effect of
  pressure on nucleation in bulk solutions and solutions in pores and
  droplets},}\ }\href@noop {} {\bibfield  {journal} {\bibinfo  {journal}
  {Journal of Colloid and Interface Science}\ }\textbf {\bibinfo {volume}
  {169}},\ \bibinfo {pages} {214--219} (\bibinfo {year} {1995})}\BibitemShut
  {NoStop}%
\bibitem [{\citenamefont {Saclier}, \citenamefont {Peczalski},\ and\
  \citenamefont {Andrieu}(2010)}]{saclier2010}%
  \BibitemOpen
  \bibfield  {author} {\bibinfo {author} {\bibfnamefont {M.}~\bibnamefont
  {Saclier}}, \bibinfo {author} {\bibfnamefont {R.}~\bibnamefont {Peczalski}},
  \ and\ \bibinfo {author} {\bibfnamefont {J.}~\bibnamefont {Andrieu}},\
  }\bibfield  {title} {\enquote {\bibinfo {title} {A theoretical model for ice
  primary nucleation induced by acoustic cavitation},}\ }\href@noop {}
  {\bibfield  {journal} {\bibinfo  {journal} {Ultrasonics Sonochemistry}\
  }\textbf {\bibinfo {volume} {17}},\ \bibinfo {pages} {98--105} (\bibinfo
  {year} {2010})}\BibitemShut {NoStop}%
\bibitem [{\citenamefont {Louisnard}, \citenamefont {Gomez},\ and\
  \citenamefont {Grossier}(2007)}]{louisnard2007}%
  \BibitemOpen
  \bibfield  {author} {\bibinfo {author} {\bibfnamefont {O.}~\bibnamefont
  {Louisnard}}, \bibinfo {author} {\bibfnamefont {F.~J.}\ \bibnamefont
  {Gomez}}, \ and\ \bibinfo {author} {\bibfnamefont {R.}~\bibnamefont
  {Grossier}},\ }\bibfield  {title} {\enquote {\bibinfo {title} {Segregation of
  a liquid mixture by a radially oscillating bubble},}\ }\href@noop {}
  {\bibfield  {journal} {\bibinfo  {journal} {Journal of Fluid Mechanics}\
  }\textbf {\bibinfo {volume} {577}},\ \bibinfo {pages} {385--415} (\bibinfo
  {year} {2007})}\BibitemShut {NoStop}%
\bibitem [{\citenamefont {Kashchiev}(2000)}]{kashchiev2000}%
  \BibitemOpen
  \bibfield  {author} {\bibinfo {author} {\bibfnamefont {D.}~\bibnamefont
  {Kashchiev}},\ }\href@noop {} {\emph {\bibinfo {title} {Nucleation}}}\
  (\bibinfo  {publisher} {Butterworth-Heinemann},\ \bibinfo {year}
  {2000})\BibitemShut {NoStop}%
\bibitem [{\citenamefont {Hirschfelder}, \citenamefont {Curtiss},\ and\
  \citenamefont {Bird}(1954)}]{hirschfelder1954}%
  \BibitemOpen
  \bibfield  {author} {\bibinfo {author} {\bibfnamefont {J.~O.}\ \bibnamefont
  {Hirschfelder}}, \bibinfo {author} {\bibfnamefont {C.~F.}\ \bibnamefont
  {Curtiss}}, \ and\ \bibinfo {author} {\bibfnamefont {R.~B.}\ \bibnamefont
  {Bird}},\ }\href@noop {} {\emph {\bibinfo {title} {Molecular theory of gases
  and liquids}}},\ Structure of Matter Series\ (\bibinfo  {publisher} {Wiley},\
  \bibinfo {year} {1954})\BibitemShut {NoStop}%
\bibitem [{\citenamefont {Bird}, \citenamefont {Stewart},\ and\ \citenamefont
  {Lightfoot}(1960)}]{bird1960}%
  \BibitemOpen
  \bibfield  {author} {\bibinfo {author} {\bibfnamefont {R.~B.}\ \bibnamefont
  {Bird}}, \bibinfo {author} {\bibfnamefont {W.~E.}\ \bibnamefont {Stewart}}, \
  and\ \bibinfo {author} {\bibfnamefont {E.~N.}\ \bibnamefont {Lightfoot}},\
  }\href@noop {} {\emph {\bibinfo {title} {Transport phenomena}}},\ Wiley
  international edition\ (\bibinfo  {publisher} {Wiley},\ \bibinfo {year}
  {1960})\BibitemShut {NoStop}%
\bibitem [{\citenamefont {Ford}(2001)}]{ford2001}%
  \BibitemOpen
  \bibfield  {author} {\bibinfo {author} {\bibfnamefont {I.~J.}\ \bibnamefont
  {Ford}},\ }\bibfield  {title} {\enquote {\bibinfo {title} {Properties of ice
  clusters from an analysis of freezing nucleation},}\ }\href@noop {}
  {\bibfield  {journal} {\bibinfo  {journal} {The Journal of Physical Chemistry
  B}\ }\textbf {\bibinfo {volume} {105}},\ \bibinfo {pages} {11649--11655}
  (\bibinfo {year} {2001})}\BibitemShut {NoStop}%
\bibitem [{\citenamefont {Kashchiev}(2006)}]{kashchiev2006}%
  \BibitemOpen
  \bibfield  {author} {\bibinfo {author} {\bibfnamefont {D.}~\bibnamefont
  {Kashchiev}},\ }\bibfield  {title} {\enquote {\bibinfo {title} {Forms and
  applications of the nucleation theorem.}}\ }\href@noop {} {\bibfield
  {journal} {\bibinfo  {journal} {The Journal of Chemical Physics}\ }\textbf
  {\bibinfo {volume} {125}},\ \bibinfo {pages} {014502} (\bibinfo {year}
  {2006})}\BibitemShut {NoStop}%
\bibitem [{\citenamefont {Ford}(1996)}]{ford1996}%
  \BibitemOpen
  \bibfield  {author} {\bibinfo {author} {\bibfnamefont {I.~J.}\ \bibnamefont
  {Ford}},\ }\bibfield  {title} {\enquote {\bibinfo {title} {Thermodynamic
  properties of critical clusters from measurements of vapour--liquid
  homogeneous nucleation rates},}\ }\href@noop {} {\bibfield  {journal}
  {\bibinfo  {journal} {The Journal of Chemical Physics}\ }\textbf {\bibinfo
  {volume} {105}},\ \bibinfo {pages} {8324--8332} (\bibinfo {year}
  {1996})}\BibitemShut {NoStop}%
\bibitem [{\citenamefont {Guggenheim}(1985)}]{guggenheim1985}%
  \BibitemOpen
  \bibfield  {author} {\bibinfo {author} {\bibfnamefont {E.~A.}\ \bibnamefont
  {Guggenheim}},\ }\bibfield  {title} {\enquote {\bibinfo {title}
  {Thermodynamics- an advanced treatment for chemists and physicists},}\
  }\href@noop {} {\bibfield  {journal} {\bibinfo  {journal} {Amsterdam,
  North-Holland, 1985, 414}\ } (\bibinfo {year} {1985})}\BibitemShut {NoStop}%
\bibitem [{\citenamefont {Abraham}(1974)}]{abraham1974}%
  \BibitemOpen
  \bibfield  {author} {\bibinfo {author} {\bibfnamefont {F.}~\bibnamefont
  {Abraham}},\ }\href@noop {} {\emph {\bibinfo {title} {Homogeneous nucleation
  theory: the pretransition theory of vapor condensation}}},\ Advances in
  Theoretical Chemistry\ (\bibinfo  {publisher} {Academic Press},\ \bibinfo
  {year} {1974})\BibitemShut {NoStop}%
\bibitem [{\citenamefont {Kalikmanov}(2012)}]{kalikmanov2012}%
  \BibitemOpen
  \bibfield  {author} {\bibinfo {author} {\bibfnamefont {V.}~\bibnamefont
  {Kalikmanov}},\ }\href@noop {} {\emph {\bibinfo {title} {Nucleation
  Theory}}},\ Lecture Notes in Physics\ (\bibinfo  {publisher} {Springer
  Netherlands},\ \bibinfo {year} {2012})\BibitemShut {NoStop}%
\bibitem [{\citenamefont {Vehkamaki}(2006)}]{vehkamaki2006}%
  \BibitemOpen
  \bibfield  {author} {\bibinfo {author} {\bibfnamefont {H.}~\bibnamefont
  {Vehkamaki}},\ }\href@noop {} {\emph {\bibinfo {title} {Classical Nucleation
  Theory in Multicomponent Systems}}}\ (\bibinfo  {publisher} {Springer},\
  \bibinfo {year} {2006})\BibitemShut {NoStop}%
\bibitem [{\citenamefont {Baidakov}, \citenamefont {Kaverin},\ and\
  \citenamefont {Boltachev}(1997)}]{baidakov1997}%
  \BibitemOpen
  \bibfield  {author} {\bibinfo {author} {\bibfnamefont {V.}~\bibnamefont
  {Baidakov}}, \bibinfo {author} {\bibfnamefont {A.}~\bibnamefont {Kaverin}}, \
  and\ \bibinfo {author} {\bibfnamefont {G.~S.}\ \bibnamefont {Boltachev}},\
  }\bibfield  {title} {\enquote {\bibinfo {title} {Nucleation in superheated
  liquid argon--krypton solutions},}\ }\href@noop {} {\bibfield  {journal}
  {\bibinfo  {journal} {The Journal of Chemical Physics}\ }\textbf {\bibinfo
  {volume} {106}},\ \bibinfo {pages} {5648--5657} (\bibinfo {year}
  {1997})}\BibitemShut {NoStop}%
\bibitem [{\citenamefont {Ford}(1997)}]{ford1997}%
  \BibitemOpen
  \bibfield  {author} {\bibinfo {author} {\bibfnamefont {I.~J.}\ \bibnamefont
  {Ford}},\ }\bibfield  {title} {\enquote {\bibinfo {title} {Nucleation
  theorems, the statistical mechanics of molecular clusters, and a revision of
  classical nucleation theory},}\ }\href@noop {} {\bibfield  {journal}
  {\bibinfo  {journal} {Physical Review E}\ }\textbf {\bibinfo {volume} {56}},\
  \bibinfo {pages} {5615} (\bibinfo {year} {1997})}\BibitemShut {NoStop}%
\bibitem [{\citenamefont {Tolman}(1949)}]{tolman1949}%
  \BibitemOpen
  \bibfield  {author} {\bibinfo {author} {\bibfnamefont {R.~C.}\ \bibnamefont
  {Tolman}},\ }\bibfield  {title} {\enquote {\bibinfo {title} {The effect of
  droplet size on surface tension},}\ }\href@noop {} {\bibfield  {journal}
  {\bibinfo  {journal} {The Journal of Chemical Physics}\ }\textbf {\bibinfo
  {volume} {17}},\ \bibinfo {pages} {333--337} (\bibinfo {year}
  {1949})}\BibitemShut {NoStop}%
\bibitem [{\citenamefont {Helfrich}(1973)}]{helfrich1973}%
  \BibitemOpen
  \bibfield  {author} {\bibinfo {author} {\bibfnamefont {W.}~\bibnamefont
  {Helfrich}},\ }\bibfield  {title} {\enquote {\bibinfo {title} {Elastic
  properties of lipid bilayers: theory and possible experiments},}\ }\href@noop
  {} {\bibfield  {journal} {\bibinfo  {journal} {Zeitschrift fur Naturforschung
  C}\ }\textbf {\bibinfo {volume} {28}},\ \bibinfo {pages} {693--703} (\bibinfo
  {year} {1973})}\BibitemShut {NoStop}%
\bibitem [{\citenamefont {Wilhelmsen}, \citenamefont {Bedeaux},\ and\
  \citenamefont {Reguera}(2015)}]{wilhelmsen2015}%
  \BibitemOpen
  \bibfield  {author} {\bibinfo {author} {\bibfnamefont {{\O}.}~\bibnamefont
  {Wilhelmsen}}, \bibinfo {author} {\bibfnamefont {D.}~\bibnamefont {Bedeaux}},
  \ and\ \bibinfo {author} {\bibfnamefont {D.}~\bibnamefont {Reguera}},\
  }\bibfield  {title} {\enquote {\bibinfo {title} {Communication: Tolman length
  and rigidity constants of water and their role in nucleation},}\ }\href@noop
  {} {\bibfield  {journal} {\bibinfo  {journal} {The Journal of Chemical
  Physics}\ }\textbf {\bibinfo {volume} {142}},\ \bibinfo {pages} {171103}
  (\bibinfo {year} {2015})}\BibitemShut {NoStop}%
\bibitem [{\citenamefont {Kashchiev}(2003)}]{kashchiev2003}%
  \BibitemOpen
  \bibfield  {author} {\bibinfo {author} {\bibfnamefont {D.}~\bibnamefont
  {Kashchiev}},\ }\bibfield  {title} {\enquote {\bibinfo {title}
  {Thermodynamically consistent description of the work to form a nucleus of
  any size},}\ }\href@noop {} {\bibfield  {journal} {\bibinfo  {journal} {The
  Journal of Chemical Physics}\ }\textbf {\bibinfo {volume} {118}},\ \bibinfo
  {pages} {1837} (\bibinfo {year} {2003})}\BibitemShut {NoStop}%
\bibitem [{\citenamefont {Vetter}\ \emph {et~al.}(2013)\citenamefont {Vetter},
  \citenamefont {Iggland}, \citenamefont {Ochsenbein}, \citenamefont
  {Hänseler},\ and\ \citenamefont {Mazzotti}}]{vetter2013}%
  \BibitemOpen
  \bibfield  {author} {\bibinfo {author} {\bibfnamefont {T.}~\bibnamefont
  {Vetter}}, \bibinfo {author} {\bibfnamefont {M.}~\bibnamefont {Iggland}},
  \bibinfo {author} {\bibfnamefont {D.~R.}\ \bibnamefont {Ochsenbein}},
  \bibinfo {author} {\bibfnamefont {F.~S.}\ \bibnamefont {Hänseler}}, \ and\
  \bibinfo {author} {\bibfnamefont {M.}~\bibnamefont {Mazzotti}},\ }\bibfield
  {title} {\enquote {\bibinfo {title} {Modeling nucleation, growth, and ostwald
  ripening in crystallization processes: A comparison between population
  balance and kinetic rate equation},}\ }\href@noop {} {\bibfield  {journal}
  {\bibinfo  {journal} {Crystal Growth and Design}\ }\textbf {\bibinfo {volume}
  {13}},\ \bibinfo {pages} {4890--4905} (\bibinfo {year} {2013})}\BibitemShut
  {NoStop}%
\bibitem [{\citenamefont {Ozkan}\ and\ \citenamefont
  {Ortoleva}(2000)}]{ozkan2000}%
  \BibitemOpen
  \bibfield  {author} {\bibinfo {author} {\bibfnamefont {G.}~\bibnamefont
  {Ozkan}}\ and\ \bibinfo {author} {\bibfnamefont {P.}~\bibnamefont
  {Ortoleva}},\ }\bibfield  {title} {\enquote {\bibinfo {title} {A mesoscopic
  model of nucleation and ostwald ripening/stepping: Application to the silica
  polymorph system},}\ }\href@noop {} {\bibfield  {journal} {\bibinfo
  {journal} {The Journal of Chemical Physics}\ }\textbf {\bibinfo {volume}
  {112}},\ \bibinfo {pages} {10510--10525} (\bibinfo {year}
  {2000})}\BibitemShut {NoStop}%
\bibitem [{\citenamefont {Ziff}, \citenamefont {McGrady},\ and\ \citenamefont
  {Meakin}(1985)}]{ziff1985}%
  \BibitemOpen
  \bibfield  {author} {\bibinfo {author} {\bibfnamefont {R.~M.}\ \bibnamefont
  {Ziff}}, \bibinfo {author} {\bibfnamefont {E.}~\bibnamefont {McGrady}}, \
  and\ \bibinfo {author} {\bibfnamefont {P.}~\bibnamefont {Meakin}},\
  }\bibfield  {title} {\enquote {\bibinfo {title} {On the validity of
  \text{Smoluchowski} equation for cluster--cluster aggregation kinetics},}\
  }\href@noop {} {\bibfield  {journal} {\bibinfo  {journal} {The Journal of
  Chemical Physics}\ }\textbf {\bibinfo {volume} {82}},\ \bibinfo {pages}
  {5269--5274} (\bibinfo {year} {1985})}\BibitemShut {NoStop}%
\bibitem [{\citenamefont {Kashchiev}(1969)}]{kashchiev1969}%
  \BibitemOpen
  \bibfield  {author} {\bibinfo {author} {\bibfnamefont {D.}~\bibnamefont
  {Kashchiev}},\ }\bibfield  {title} {\enquote {\bibinfo {title} {Nucleation at
  variable supersaturation},}\ }\href@noop {} {\bibfield  {journal} {\bibinfo
  {journal} {Surface Science}\ }\textbf {\bibinfo {volume} {18}},\ \bibinfo
  {pages} {293--297} (\bibinfo {year} {1969})}\BibitemShut {NoStop}%
\bibitem [{\citenamefont {Chang}\ and\ \citenamefont
  {Cooper}(1970)}]{chang1970}%
  \BibitemOpen
  \bibfield  {author} {\bibinfo {author} {\bibfnamefont {J.}~\bibnamefont
  {Chang}}\ and\ \bibinfo {author} {\bibfnamefont {G.}~\bibnamefont {Cooper}},\
  }\bibfield  {title} {\enquote {\bibinfo {title} {A practical difference
  scheme for \text{Fokker-Planck} equations},}\ }\href@noop {} {\bibfield
  {journal} {\bibinfo  {journal} {Journal of Computational Physics}\ }\textbf
  {\bibinfo {volume} {6}},\ \bibinfo {pages} {1--16} (\bibinfo {year}
  {1970})}\BibitemShut {NoStop}%
\bibitem [{\citenamefont {Brown}, \citenamefont {Byrne},\ and\ \citenamefont
  {Hindmarsh}(1989)}]{brown1989}%
  \BibitemOpen
  \bibfield  {author} {\bibinfo {author} {\bibfnamefont {P.~N.}\ \bibnamefont
  {Brown}}, \bibinfo {author} {\bibfnamefont {G.~D.}\ \bibnamefont {Byrne}}, \
  and\ \bibinfo {author} {\bibfnamefont {A.~C.}\ \bibnamefont {Hindmarsh}},\
  }\bibfield  {title} {\enquote {\bibinfo {title} {\text{VODE}: A
  variable-coefficient \text{ODE} solver},}\ }\href@noop {} {\bibfield
  {journal} {\bibinfo  {journal} {SIAM Journal on Scientific and Statistical
  Computing}\ }\textbf {\bibinfo {volume} {10}},\ \bibinfo {pages} {1038--1051}
  (\bibinfo {year} {1989})}\BibitemShut {NoStop}%
\bibitem [{\citenamefont {Koz\'isek}\ and\ \citenamefont
  {Demo}(2005)}]{kozisek2005}%
  \BibitemOpen
  \bibfield  {author} {\bibinfo {author} {\bibfnamefont {Z.}~\bibnamefont
  {Koz\'isek}}\ and\ \bibinfo {author} {\bibfnamefont {P.}~\bibnamefont
  {Demo}},\ }\bibfield  {title} {\enquote {\bibinfo {title} {Influence of
  initial conditions on homogeneous nucleation kinetics in a closed system},}\
  }\href@noop {} {\bibfield  {journal} {\bibinfo  {journal} {The Journal of
  Chemical Physics}\ }\textbf {\bibinfo {volume} {123}},\ \bibinfo {pages}
  {144502} (\bibinfo {year} {2005})}\BibitemShut {NoStop}%
\bibitem [{\citenamefont {Samsonov}, \citenamefont {Bazulev},\ and\
  \citenamefont {Sdobnyakov}(2003)}]{samsonov2003}%
  \BibitemOpen
  \bibfield  {author} {\bibinfo {author} {\bibfnamefont {V.}~\bibnamefont
  {Samsonov}}, \bibinfo {author} {\bibfnamefont {A.}~\bibnamefont {Bazulev}}, \
  and\ \bibinfo {author} {\bibfnamefont {N.~Y.}\ \bibnamefont {Sdobnyakov}},\
  }\bibfield  {title} {\enquote {\bibinfo {title} {Rusanov's linear formula for
  the surface tension of small objects},}\ }in\ \href@noop {} {\emph {\bibinfo
  {booktitle} {Doklady Physical Chemistry}}},\ Vol.\ \bibinfo {volume} {389}\
  (\bibinfo {organization} {Springer},\ \bibinfo {year} {2003})\ pp.\ \bibinfo
  {pages} {83--85}\BibitemShut {NoStop}%
\bibitem [{\citenamefont {Lau}\ \emph {et~al.}(2015)\citenamefont {Lau},
  \citenamefont {Hunt}, \citenamefont {M{\"u}ller}, \citenamefont {Jackson},\
  and\ \citenamefont {Ford}}]{lau2015}%
  \BibitemOpen
  \bibfield  {author} {\bibinfo {author} {\bibfnamefont {G.~V.}\ \bibnamefont
  {Lau}}, \bibinfo {author} {\bibfnamefont {P.~A.}\ \bibnamefont {Hunt}},
  \bibinfo {author} {\bibfnamefont {E.~A.}\ \bibnamefont {M{\"u}ller}},
  \bibinfo {author} {\bibfnamefont {G.}~\bibnamefont {Jackson}}, \ and\
  \bibinfo {author} {\bibfnamefont {I.~J.}\ \bibnamefont {Ford}},\ }\bibfield
  {title} {\enquote {\bibinfo {title} {Water droplet excess free energy
  determined by cluster mitosis using guided molecular dynamics},}\ }\href@noop
  {} {\bibfield  {journal} {\bibinfo  {journal} {The Journal of Chemical
  Physics}\ }\textbf {\bibinfo {volume} {143}},\ \bibinfo {pages} {244709}
  (\bibinfo {year} {2015})}\BibitemShut {NoStop}%
\bibitem [{\citenamefont {Brus}, \citenamefont {Zdimal},\ and\ \citenamefont
  {Smolik}(2008)}]{brus2008}%
  \BibitemOpen
  \bibfield  {author} {\bibinfo {author} {\bibfnamefont {D.}~\bibnamefont
  {Brus}}, \bibinfo {author} {\bibfnamefont {V.}~\bibnamefont {Zdimal}}, \ and\
  \bibinfo {author} {\bibfnamefont {J.}~\bibnamefont {Smolik}},\ }\bibfield
  {title} {\enquote {\bibinfo {title} {Homogeneous nucleation rate measurements
  in supersaturated water vapor},}\ }\href@noop {} {\bibfield  {journal}
  {\bibinfo  {journal} {The Journal of Chemical Physics}\ }\textbf {\bibinfo
  {volume} {129}},\ \bibinfo {pages} {174501} (\bibinfo {year}
  {2008})}\BibitemShut {NoStop}%
\bibitem [{\citenamefont {Brus}, \citenamefont {Zdimal},\ and\ \citenamefont
  {Uchtmann}(2009)}]{brus2009}%
  \BibitemOpen
  \bibfield  {author} {\bibinfo {author} {\bibfnamefont {D.}~\bibnamefont
  {Brus}}, \bibinfo {author} {\bibfnamefont {V.}~\bibnamefont {Zdimal}}, \ and\
  \bibinfo {author} {\bibfnamefont {H.}~\bibnamefont {Uchtmann}},\ }\bibfield
  {title} {\enquote {\bibinfo {title} {Homogeneous nucleation rate measurements
  in supersaturated water vapor \uppercase{II}},}\ }\href@noop {} {\bibfield
  {journal} {\bibinfo  {journal} {The Journal of Chemical Physics}\ }\textbf
  {\bibinfo {volume} {131}},\ \bibinfo {pages} {074507} (\bibinfo {year}
  {2009})}\BibitemShut {NoStop}%
\bibitem [{\citenamefont {Talanquer}\ and\ \citenamefont
  {Oxtoby}(1995)}]{talanquer1995}%
  \BibitemOpen
  \bibfield  {author} {\bibinfo {author} {\bibfnamefont {V.}~\bibnamefont
  {Talanquer}}\ and\ \bibinfo {author} {\bibfnamefont {D.}~\bibnamefont
  {Oxtoby}},\ }\bibfield  {title} {\enquote {\bibinfo {title} {Density
  functional analysis of phenomenological theories of gas-liquid nucleation},}\
  }\href@noop {} {\bibfield  {journal} {\bibinfo  {journal} {The Journal of
  Physical Chemistry}\ }\textbf {\bibinfo {volume} {99}},\ \bibinfo {pages}
  {2865--2874} (\bibinfo {year} {1995})}\BibitemShut {NoStop}%
\bibitem [{\citenamefont {Lu}\ and\ \citenamefont {Jiang}(2005)}]{lu2005}%
  \BibitemOpen
  \bibfield  {author} {\bibinfo {author} {\bibfnamefont {H.~M.}\ \bibnamefont
  {Lu}}\ and\ \bibinfo {author} {\bibfnamefont {Q.}~\bibnamefont {Jiang}},\
  }\bibfield  {title} {\enquote {\bibinfo {title} {Size-dependent surface
  tension and \text{Tolman's} length of droplets},}\ }\href@noop {} {\bibfield
  {journal} {\bibinfo  {journal} {Langmuir}\ }\textbf {\bibinfo {volume}
  {21}},\ \bibinfo {pages} {779--781} (\bibinfo {year} {2005})}\BibitemShut
  {NoStop}%
\bibitem [{\citenamefont {Granasy}(1998)}]{granasy1998}%
  \BibitemOpen
  \bibfield  {author} {\bibinfo {author} {\bibfnamefont {L.}~\bibnamefont
  {Granasy}},\ }\bibfield  {title} {\enquote {\bibinfo {title} {Semiempirical
  van der \text{Waals}/\text{Cahn-Hilliard} theory: Size dependence of the
  \text{Tolman's} length},}\ }\href@noop {} {\bibfield  {journal} {\bibinfo
  {journal} {The Journal of Chemical Physics}\ }\textbf {\bibinfo {volume}
  {109}},\ \bibinfo {pages} {9660--9663} (\bibinfo {year} {1998})}\BibitemShut
  {NoStop}%
\bibitem [{\citenamefont {W{\"o}lk}\ and\ \citenamefont
  {Strey}(2001)}]{wolk2001}%
  \BibitemOpen
  \bibfield  {author} {\bibinfo {author} {\bibfnamefont {J.}~\bibnamefont
  {W{\"o}lk}}\ and\ \bibinfo {author} {\bibfnamefont {R.}~\bibnamefont
  {Strey}},\ }\bibfield  {title} {\enquote {\bibinfo {title} {Homogeneous
  nucleation of \text{H2O and D2O} in comparison: the isotope effect},}\
  }\href@noop {} {\bibfield  {journal} {\bibinfo  {journal} {The Journal of
  Physical Chemistry B}\ }\textbf {\bibinfo {volume} {105}},\ \bibinfo {pages}
  {11683--11701} (\bibinfo {year} {2001})}\BibitemShut {NoStop}%
\bibitem [{\citenamefont {Holten}, \citenamefont {Labetski},\ and\
  \citenamefont {van Dongen}(2005)}]{holten2005}%
  \BibitemOpen
  \bibfield  {author} {\bibinfo {author} {\bibfnamefont {V.}~\bibnamefont
  {Holten}}, \bibinfo {author} {\bibfnamefont {D.~G.}\ \bibnamefont
  {Labetski}}, \ and\ \bibinfo {author} {\bibfnamefont {M.~E.~H.}\ \bibnamefont
  {van Dongen}},\ }\bibfield  {title} {\enquote {\bibinfo {title} {Homogeneous
  nucleation of water between 200 and 240 k: new wave tube data and estimation
  of the \text{Tolman} length},}\ }\href@noop {} {\bibfield  {journal}
  {\bibinfo  {journal} {The Journal of Chemical Physics}\ }\textbf {\bibinfo
  {volume} {123}},\ \bibinfo {pages} {104505} (\bibinfo {year}
  {2005})}\BibitemShut {NoStop}%
\bibitem [{\citenamefont {Baidakov}\ and\ \citenamefont
  {Kaverin}(2009)}]{baidakov2009}%
  \BibitemOpen
  \bibfield  {author} {\bibinfo {author} {\bibfnamefont {V.~G.}\ \bibnamefont
  {Baidakov}}\ and\ \bibinfo {author} {\bibfnamefont {A.~M.}\ \bibnamefont
  {Kaverin}},\ }\bibfield  {title} {\enquote {\bibinfo {title} {Boiling-up of
  superheated liquid argon in an acoustic field},}\ }\href@noop {} {\bibfield
  {journal} {\bibinfo  {journal} {Journal of Physics: Condensed Matter}\
  }\textbf {\bibinfo {volume} {21}},\ \bibinfo {pages} {465103} (\bibinfo
  {year} {2009})}\BibitemShut {NoStop}%
\bibitem [{\citenamefont {Canney}\ \emph {et~al.}(2008)\citenamefont {Canney},
  \citenamefont {Bailey}, \citenamefont {Crum}, \citenamefont {Khokhlova},\
  and\ \citenamefont {Sapozhnikov}}]{crum2008}%
  \BibitemOpen
  \bibfield  {author} {\bibinfo {author} {\bibfnamefont {M.~S.}\ \bibnamefont
  {Canney}}, \bibinfo {author} {\bibfnamefont {M.~R.}\ \bibnamefont {Bailey}},
  \bibinfo {author} {\bibfnamefont {L.~A.}\ \bibnamefont {Crum}}, \bibinfo
  {author} {\bibfnamefont {V.~A.}\ \bibnamefont {Khokhlova}}, \ and\ \bibinfo
  {author} {\bibfnamefont {O.~A.}\ \bibnamefont {Sapozhnikov}},\ }\bibfield
  {title} {\enquote {\bibinfo {title} {Acoustic characterization of high
  intensity focused ultrasound fields: A combined measurement and modeling
  approach},}\ }\href@noop {} {\bibfield  {journal} {\bibinfo  {journal} {The
  Journal of the Acoustical Society of America}\ }\textbf {\bibinfo {volume}
  {124}},\ \bibinfo {pages} {2406--2420} (\bibinfo {year} {2008})}\BibitemShut
  {NoStop}%
\bibitem [{\citenamefont {Harzali}\ \emph {et~al.}(2011)\citenamefont
  {Harzali}, \citenamefont {Baillon}, \citenamefont {Louisnard}, \citenamefont
  {Espitalier},\ and\ \citenamefont {Mgaidi}}]{harzali2011}%
  \BibitemOpen
  \bibfield  {author} {\bibinfo {author} {\bibfnamefont {H.}~\bibnamefont
  {Harzali}}, \bibinfo {author} {\bibfnamefont {F.}~\bibnamefont {Baillon}},
  \bibinfo {author} {\bibfnamefont {O.}~\bibnamefont {Louisnard}}, \bibinfo
  {author} {\bibfnamefont {F.}~\bibnamefont {Espitalier}}, \ and\ \bibinfo
  {author} {\bibfnamefont {A.}~\bibnamefont {Mgaidi}},\ }\bibfield  {title}
  {\enquote {\bibinfo {title} {Experimental study of sono-crystallisation of
  \text{ZnSO4}{\textperiodcentered} \text{7H2O}, and interpretation by the
  segregation theory},}\ }\href@noop {} {\bibfield  {journal} {\bibinfo
  {journal} {Ultrasonics Sonochemistry}\ }\textbf {\bibinfo {volume} {18}},\
  \bibinfo {pages} {1097--1106} (\bibinfo {year} {2011})}\BibitemShut {NoStop}%
\end{thebibliography}

%

\end{document}